\documentstyle[11pt,epsfig,aaspp4]{article}
\tighten 
\begin{document}

\newcommand{\lya}{Lyman~$\alpha$}
\newcommand{\lyb}{Lyman~$\beta$}
\newcommand{\za}{$z_{\rm abs}$}
\newcommand{\ze}{$z_{\rm em}$}
\newcommand{\cmtwo}{cm$^{-2}$}
\newcommand{\nhi}{$N$(H$^0$)}
\newcommand{\nzn}{$N$(Zn$^+$)}
\newcommand{\ncr}{$N$(Cr$^+$)}
\newcommand{\degpoint}{\mbox{$^\circ\mskip-7.0mu.\,$}}
\newcommand{\halpha}{\mbox{H$\alpha$}}
\newcommand{\hbeta}{\mbox{H$\beta$}}
\newcommand{\hgamma}{\mbox{H$\gamma$}}
\newcommand{\kms}{\,km~s$^{-1}$}      
\newcommand{\minpoint}{\mbox{$'\mskip-4.7mu.\mskip0.8mu$}}
\newcommand{\mv}{\mbox{$m_{_V}$}}
\newcommand{\Mv}{\mbox{$M_{_V}$}}
\newcommand{\peryr}{\mbox{$\>\rm yr^{-1}$}}
\newcommand{\secpoint}{\mbox{$''\mskip-7.6mu.\,$}}
\newcommand{\sqdeg}{\mbox{${\rm deg}^2$}}
\newcommand{\squig}{\sim\!\!}
\newcommand{\subsun}{\mbox{$_{\twelvesy\odot}$}}
\newcommand{\et}{et al.~}

\def\ltsima{$\; \buildrel < \over \sim \;$}
\def\simlt{\lower.5ex\hbox{\ltsima}}
\def\gtsima{$\; \buildrel > \over \sim \;$}
\def\simgt{\lower.5ex\hbox{\gtsima}}
\def\arcs{$''~$}
\def\arcm{$'~$}
\vspace*{0.1cm}
\title{NEW OBSERVATIONS OF THE INTERSTELLAR MEDIUM 
IN THE LYMAN BREAK GALAXY MS~1512$-$cB58
\altaffilmark{1}}

\vspace{1cm}
\author{\sc Max Pettini and Samantha A. Rix}
\affil{Institute of Astronomy, Madingley Road, Cambridge, CB3 0HA, UK}
\author{\sc Charles C. Steidel\altaffilmark{2}}
\affil{Palomar Observatory, Caltech 105--24, Pasadena, CA 91125}
\author{\sc Kurt L. Adelberger\altaffilmark{3}}
\affil{Center for Astrophysics, 60 Garden Street, Cambridge, MA 02138}
\author{\sc Matthew P. Hunt and Alice E. Shapley}
\affil{Palomar Observatory, Caltech 105--24, Pasadena, CA 91125}

\altaffiltext{1}{Based on data obtained at the W. M. Keck Observatory which
is operated as a scientific partnership among the California 
Institute of Technology, the
University of California, and NASA, and was made possible by
the generous financial support of the W.M. Keck Foundation. }
\altaffiltext{2}{Packard Fellow}
\altaffiltext{3}{Harvard Society Junior Fellow} 

\newpage
\begin{abstract}
We present the results of a detailed study of the interstellar medium
(ISM) of MS~1512$-$cB58, an $\sim L^{\ast}$ Lyman break galaxy
at $z = 2.7276$, based on new spectral observations obtained with
the Echelle Spectrograph and Imager on the Keck~II telescope
at 58\kms\ resolution.
We focus in particular on the chemical abundances
and kinematics of the interstellar gas deduced from
the analysis of 48 ultraviolet absorption lines, 
at rest wavelengths between 1134 and 2576\,\AA,
due to elements from H to Zn. Our main findings are as follows.
Even at this relatively early epoch, the ISM of this
galaxy is already highly enriched in elements
released by Type~II supernovae; the abundances of O, Mg, Si, P, and S
are all $\sim 2/5$ of their solar values. In contrast, N and the Fe-peak
elements Mn, Fe, and Ni are underabundant by a factor of $\sim 3$.
Based on current ideas of stellar nucleosynthesis, these results
can be understood if most of the metal enrichment in cB58 has taken
place within the last $\sim 300$\,Myr, the timescale for the
release of N from intermediate mass stars. Such a young age is consistent
with the UV-optical spectral energy distribution. Thus,
cB58 appears to be an example of a galaxy in the process
of converting its gas into stars on a few dynamical timescales---quite
possibly we are witnessing the formation of a galactic bulge
or an elliptical galaxy.
The energetic star formation activity has stirred the 
interstellar medium to high velocities; the strongest absorption lines
span a velocity interval of $\sim 1000$\,\kms.  
The net effect is a bulk outflow of the 
ISM at a speed of $\sim 255$\,\kms\ and at a rate which exceeds
the star formation rate. 
It is unclear whether this gas will be lost or retained 
by the galaxy. On the one hand, the outflow probably has
sufficient energy to escape the potential well
of cB58, for which we derive a baryonic mass of 
$\sim 10^{10} M_{\odot}$.
On the other hand, at least some of the 
elements manufactured
by previous generations of stars must have mixed
efficiently with the ambient, neutral, ISM to give
the high abundances we measure.
We point out that the chemical and kinematic 
properties of cB58 are markedly different from those of 
most damped \lya\ systems at the same redshift.
\end{abstract}
\keywords{cosmology: observations --- galaxies: evolution ---
galaxies: starburst --- galaxies: individual (MS~1512$-$cB58)}

\newpage
\section{INTRODUCTION}
Thanks to its gravitationally lensed nature,
the $z = 2.7276$ galaxy MS~1512$-$cB58
(or cB58 for short)
provides an unusually clear window on the population
of star-forming galaxies identified at these redshifts through
the Lyman break technique (Steidel et al. 1996).
Discovered by Yee et al. (1996), 
cB58 is, as far as we can tell, a
typical $\sim L^{\ast}$ galaxy 
magnified by a factor of $\sim 30$
by the foreground cluster MS~1512$+$36 at $z = 0.37$
(Seitz et al. 1998). This fortuitous alignment
makes it by far the brightest known member of the
Lyman break galaxy (LBG) population and has motivated
a number of studies 
from mm to X-ray wavelengths
(Frayer et al. 1997; Nakanishi et al. 1997;
Baker et al. 2001; Sawicki 2001;
Teplitz et al. 2000; Pettini et al. 2000a;
Almaini et al., in preparation).
In Pettini et al. (2000a, hereinafter Paper I) 
we described the 
main properties of the rest-frame ultraviolet
spectrum of the galaxy; here we follow on that work
with a more in-depth analysis made possible
by the superior quality and wider wavelength coverage 
of recent spectroscopic
observations with the Echelle Spectrograph and Imager
(ESI) on the Keck~II telescope. We focus in particular
on the interstellar lines, which are fully resolved
in the new data, with two main objectives.
First, we consider the kinematics of the 
interstellar medium which show evidence for 
large scale outflows of gas and dust; such `superwinds'
are commonly observed in starburst galaxies at low and
high redshifts (Heckman et al. 2000; Pettini et al. 2001),
but have so far been little studied 
in the ultraviolet (UV) regime.
Second, we shall attempt to deduce a comprehensive 
picture of the chemical composition of the interstellar gas,
with the attendant clues which such information may
provide on the evolutionary status of the galaxy and its 
past history of star formation.
Abundance data for LBGs are still very patchy
(Pettini 1999; Pettini et al. 2001 and references 
therein); apart from the opportunity to 
determine element abundances with 
some degree of confidence for the interstellar gas
in cB58, we also hold the hope that 
the study of this unusually bright object may eventually
lead us to identify some UV abundance indicators which 
can be applied more generally to the whole LBG population.\\

\section{OBSERVATIONS AND DATA REDUCTION}

ESI (Sheinis et al. 2000) is an efficient new instrument 
on the Keck~II telescope; in its echelle mode it 
covers in a single $2048\times4096$ 15$\mu$m pixel
CCD frame the whole wavelength range from 4000 to 10\,500\,\AA\
with 10 spectral orders at a dispersion of 
$\sim 11.5$\,km~s$^{-1}$ per pixel. 
The current detector is an MIT/Lincoln Labs CCDID20 CCD
which has a quantum efficiency greater than 70\% from 4500 to 
nearly 8000\,\AA.

We observed cB58 with ESI on the nights of May 30, May 31 and June 1, 
2000 for a total of 16\,000\,s ($8 \times 2\,000$\,s exposures)
in good conditions, with seeing between 0.65\,\arcs\ and 0.85\,\arcs.
We used a 1\,\arcs wide entrance slit aligned at 
position angle PA = 93$^{\circ}$, along the
long axis of cB58 which is distorted into a gravitational fold arc 
approximately 3\,\arcs long. 
All the observations were conducted at  
airmass between 1.04 and 1.08\,.

The ESI images were reduced, coadded and 
mapped onto a common vacuum heliocentric wavelength scale
with the ``Dukee'' IRAF reduction package written by one us
(KLA). Even for an extended object such as cB58,
the minimum 20\,\arcs separation between adjacent
echelle orders provides adequate coverage of the 
sky background. Consequently, subtraction of
the sky signal did not degrade significantly
the signal-to-noise ratio and produced no obvious
systematic error---that is, in the black cores of the
strongest interstellar absorption lines we measured a 
mean count consistent with zero within the 
random noise. The one exception was the region near
the damped \lya\ line in cB58 which,
at $\lambda_{\rm obs} = 4527$\,\AA,
happens to fall close to the only serious CCD defect
in the whole detector area. Here we applied a
positive correction (that is, we {\it added}
counts to raise the core of the line from negative flux
values to zero), determined empirically and amounting 
to $\sim 7\%$ of the continuum level.
Flux calibration was by reference to the 
spectra of the spectrophotometric standard stars
Feige 34 and Feige 110 recorded on each night.

The final spectrum has a signal-to-noise ratio per resolution 
element of between $\sim 30$  and $\sim 55$
(the S/N varies along each echelle order)
from $\sim 4850$ to $8750$\,\AA\ ($\sim 1300-2350$\,\AA\
in the rest-frame of cB58). From the widths of the 
night sky emission lines we measured a  
spectral resolution of 58\,km~s$^{-1}$ FWHM.\\

\section{THE INTERSTELLAR SPECTRUM OF MS~1512$-$cB58}

\subsection{The Systemic Redshift of the Galaxy}

The ESI spectrum of cB58 exhibits a multitude
of stellar and interstellar lines, as well as absorption
from the \lya\ forest and several intervening
metal line systems. Most of the stellar
lines are blends from multiple transitions.
The clearest single photospheric line is C~III~$\lambda 2296.871$
(laboratory wavelength in air, corresponding
to vacuum wavelength $\lambda_{\rm vac} = 2297.579$);
we observe this line at a central wavelength 
$\lambda_{\rm obs} = 8564.4^{+0.2}_{-0.6}$\,\AA\
from which we deduce a redshift 
$z_{\rm stars} = 2.7276^{+0.0001}_{-0.0003}$
(the asymmetric errors arise from a weak feature
in the blue wing which may or may not
be part of the line)\footnote{All redshifts quoted
in this paper are vacuum heliocentric.}.
Within the errors, this value 
is consistent with $z_{\rm stars} = 2.7268 \pm 0.0008$
derived in Paper I from a number of less well defined 
photospheric lines in the far-UV.
Teplitz et al. (2000) measured 
$z_{\rm H II} = 2.7290 \pm 0.0007$
from the mean of ten emission lines
formed in H~II regions---from [O~II]~$\lambda 3727$
to [N II]~$\lambda 6583$---redshifted into
the near-infrared. It is unclear whether
the $2 \sigma$ difference 
between $z_{\rm H II}$ and $z_{\rm stars}$
reflects a real velocity offset of $\sim 100$\,km~s$^{-1}$
between the stars and the ionized gas,
or is due to a systematic difference
between the wavelength calibrations
of the optical and infrared observations.
In any case, here we adopt $z_{\rm stars} = 2.7276$
as the systemic redshift of cB58.\\

\subsection{Interstellar Absorption Lines}

Table 1 lists the interstellar absorption lines
measured in our ESI spectrum of cB58.
Together with \lya\ (not included in Table 1
but discussed separately in $\S 4.1$) 
we cover 48 transitions of elements
from H to Zn in a variety of ionization
stages, from neutral (H~I, C~I, O~I, N~I)
to highly ionized species (Si~IV, C~IV, N~V).
Vacuum rest wavelengths and $f$-values of the
transitions are from the compilation
by Morton (1991) updated with subsequent revisions
as listed by Savage \& Sembach (1996) and
R.F. Carswell (private communication).
The interstellar lines are seen against the continuum
provided by the integrated light of
O and B stars in the galaxy.
The resulting composite stellar spectrum 
is rich in photospheric and wind lines
which are the subject of a future paper (see also Paper I).
The resolution of the ESI data is adequate to
allow, with some care, stellar and interstellar features to 
be separated in most cases; in this we are aided by the fact
that the strongest interstellar absorption is centered at
$-255$\,\kms\ relative to the stellar redshift (\S3.3).
However, many of the interstellar
lines at wavelengths below 1216\,\AA\ could not be recovered
because they are blended
with intervening \lya\ forest lines.

Rest-frame equivalent widths, $W_0$, are listed in column (6)
of Table 1 together with their $1 \sigma$ errors in
column (7); they were measured by summing
the absorption over fixed velocity ranges
$\Delta v$ indicated in column (5) of Table 1. 
The velocity ranges were chosen to encompass
the full extent of the absorption while minimizing
the amount of continuum included (and therefore
the error on the equivalent width measurement); 
thus values of $\Delta v$ are larger for the stronger lines.
In a few cases (mostly in the \lya\ forest)
where an interstellar line 
was found to be blended with
other unrelated features over the interval $\Delta v$, 
we list its equivalent
width as a lower limit
which does not include the blend.
Figures 1 and 2 are a compendium of absorption
lines from different ionization stages
of a variety of elements.
We now describe the velocity structure
of the absorbing gas.

\subsection{Kinematics of the Absorbing Gas}

The interstellar medium of cB58 has obviously been stirred
and accelerated to high velocities, presumably
by the kinetic energy deposited by the star formation
activity via stellar winds and supernovae.
The strongest transitions in Figures 1 and 2
show absorption over $\Delta v \sim 1000$\kms, from
$\sim -775$ to $\sim +225$\kms\ relative to the 
redshift of the stars $z_{\rm stars} = 2.7276$\,.
Gas with the largest optical depth occurs at velocities near
$v_{\rm ISM} \simeq -255$\kms, corresponding 
to an absorption redshift $z_{\rm abs} = 2.7244$.
Thus the overall effect we are seeing is an outflow
of the ISM from the star formation region.
The outflow speed of $-255$\kms\ is typical of 
Lyman break galaxies.
It is now well established that
the {\it centroids} of the interstellar lines
are normally blueshifted by between $\sim 200$
and $\sim 400$\kms\ relative 
to the emission lines from H~II regions, at least
in LBGs brighter than $L^{\ast}$
(Pettini et al. 2001 and references therein).
What the new observations presented here reveal
is that there is gas in front of the stars
with a much wider range of velocities,
from $\sim -500$ to $\sim +500$\kms,
relative to the bulk of the outflow.
We have no reason to believe that such extreme
velocities are not commonly encountered in 
a significant fraction of LBGs, at least those whose
spectra show strong interstellar absorption 
lines. At the low spectral resolution 
of most previous studies the velocity structure
so evident here cannot be discerned,
although the line equivalent widths themselves 
are indicative of values of $\Delta v$ of 
several hundred \kms\ (Steidel et al. 1996).

Inspection of Figures 1 and 2
shows that all the ion stages observed
have similar absorption profiles.
First, they span the same overall velocity range,
as can be realized by comparing, for example,
Si~II~$\lambda 1526$ in Figure 1 with 
the Si~III and Si~IV lines 
in Figure 2. 
Second, the gas with the highest optical depth 
is found at very similar velocities 
near $-255$\kms; there is a shift of only 
$\sim 20$\kms\ between the neutrals and first ions
($v_{\tau~{\rm max}} \simeq -265$\kms; see Figure 1)
and highly ionized species from Al~III to N~V
($v_{\tau~{\rm max}} \simeq -245$\kms; see Figure 2).
The main difference between species which are 
the dominant ions in  H~I regions
and those which trace ionized gas is that the 
latter show smoother absorption profiles.
The Al~III, Si~IV and C~IV doublets in Figure 2
are remarkable in their almost continuous
decrease in optical depth over the interval
$\pm 500$\kms\ from the central velocity
of $-245$\kms, while the first ions in Figure 1
break up into a number of 
seemingly discrete components. 

The spatial and kinematic structure of the interstellar medium
in cB58 is clearly very complex. 
Given the lack of detailed information, we can only speculate 
on the location and nature of the  
different absorption components.
In the simplest picture, the starburst
produces a large bubble of hot gas which
sweeps out ambient interstellar matter as it expands
and the main component of the absorption 
arises in compressed, cooling gas behind the shock front.
Such a simple superbubble model can account for many
(although not all) of the properties of the outflow
seen in the local starburst galaxy 
NGC~1705 (Heckman et al. 2001b),
but encounters some difficulties in the case of cB58.
First, the cooling would have to be extremely efficient
to give the high column densities of neutral gas we see (\S4.1),
since the high velocities imply shock temperatures 
$T = 2 \times 10^6 \times (v/300\,{\rm km~s^{-1}})^2$~K,
in the X-ray regime (Wallerstein \& Silk 1971).
Second, while the main absorption component centered
near $-255$\,\kms, which as we shall see below 
(\S 4.2) accounts for approximately two thirds of the column
density of neutral gas, could in principle arise in
such a supershell, the simple model provides no
explanation for the rest of the absorption with
$|\Delta v|$ of up to 500\,\kms\ relative to it.
Finally, a velocity of $-255$\,\kms\ for the swept-up
shell of interstellar matter appears rather high.
The expansion speed of the 
shell is related to the rate of deposition
of mechanical energy by the starburst, $\dot{E}_{42}$
(in units of $10^{42}$\,erg s$^{-1}$), by the expression (e.g. Heckman 2001):
\begin{equation}
v_{\rm shell}\, \sim 100 \, \dot{E}_{42}^{1/5} \, n_0^{-1/5} \, t_7^{-2/5} \,  {\rm km~s^{-1}}
\end{equation}
where $n_0$ is the ambient density and $t_7$ is the time since the expansion
began in units of $10^7$\, years.
From {\it Starburst99} (Leitherer et al. 1999) we find that 
the UV luminosity of cB58, $L_{1500} = 8 \times 10^{29}$\, erg~s$^{-1}$~Hz$^{-1}$
(corrected for gravitational lensing and dust extinction; 
see Paper I and references therein),
corresponds to $\dot{E}_{42} = 32$. 
The spectral energy distribution of cB58 has been interpreted
by Ellingson et al. (1996) and Bechtold et al. (1998)
as evidence for an age $t_7 \simlt 3.5$.
From these parameters we would expect 
$v_{\rm shell} \simeq 120$\,\kms\ for unity density
of the ambient ISM.  In order to reproduce the observed
$v \simeq -255$\,\kms\ we require either 
$n_0 \ll 1$ or $t_7 < 1$; while the former is 
a possibility, the latter seems unlikely because it would
imply that we are viewing cB58 at a special time, soon
after the onset of the starburst.

In an alternative 
scenario (see for example Strickland \& Stevens 2000), 
the absorption arises {\it within} the cavity
(filled with $T \sim 10^7$\,K gas which is hard to observe
directly) from fragments of the 
ambient ISM ablated by the hot gas. As these fragments
are entrained in the flow they are accelerated 
from rest (relative to the stars) to speeds of up to
$\sim 750$\kms\ before they are heated and ionized
to levels which are no longer observable with
UV absorption line spectroscopy.
While physically more plausible, this scenario
does not naturally account for (a) the fact that
most of the neutral gas is moving at
$v = -265$\,\kms\ relative to the stars, and 
(b) the similarity in the velocity profiles
of absorption lines from different ionization stages,
from species which are most abundant 
in neutral gas to the highly ionized Si~IV and C~IV.
Naively, if the absorption arises from mass-loading
of the flow, we would expect to see a progression 
towards more negative velocities with increasing ionization, 
as material is entrained and ionized by the ouflowing hot gas.
This is the case in NGC~1705 (Heckman et al. 2001b).
Possibly, such velocity differences are smeared
and made difficult to recognize by the fact that
we see absorption against an 
extended stellar background (see below), so that
our spectrum is in fact the sum of many 
sightlines through the galaxy.

In any case, since the hydrodynamics of galactic superwinds are
still poorly constrained, it is worthwhile pointing out here
a few empirical considerations which should be accounted for in
future theoretical treatments. First, there appears to be a high degree
of symmetry in the kinematics of the outflow in cB58. Not only
are the absorption profiles of different ionization stages
broadly similar, as explained above, but the receding 
part of the outflow, which lies {\it behind} the stars
and is detected via \lya\ emission (\S4.1)
also appears to exhibit essentially the same kinematics,
mirroring at positive velocities the pattern seen in absorption.
Given the stochastic nature of \lya\ emission, however,
this could be no more than a coincidence---in other LBGs
the peak of \lya\ emission is often at {\it higher} positive
velocities than the blueshift of the absorption lines
(see Figure 13 of Pettini et al. 2001).
Second, whatever its location, 
the bulk of the gas moving at 
$v = -255$\kms\ must be located well in front of the stars
and cover them completely since it absorbs all of the UV light;
as can be seen from Figures 1 and 2, 
the strongest metal absorption lines
of all ion stages---from O~I to C~IV---are 
black at this velocity.
From their detailed analysis
of WFPC2 $R$ and $V$ images, Seitz et al. (1998)
derived a source size $r_{1/2} = 0.25 \pm 0.05$\,\arcs
(where $r_{1/2}$ is the half-light radius)
which is magnified and distorted by gravitational
lensing into a gravitational fold arc about 
3\,\arcs long and with an axis ratio of 1:7.
A half-light radius of 0.25\,\arcs corresponds 
to linear dimensions
$r_{1/2} = 2 h_{70}^{-1}$\,kpc
for an $\Omega_M =0.3$, $\Omega_{\Lambda} = 0.7$
cosmology; the structure responsible
for the bulk of the absorption must therefore extend 
more than $2r_{1/2} =4 h_{70}^{-1}$\,kpc across the line of sight.
We return to the properties of the outflow in \S6.3.\\

\section{ION COLUMN DENSITIES}

\subsection{The Lyman $\alpha$ Line}

As can be seen from Figure 3, 
the \lya\ line in cB58 is a combination of absorption and emission.
The new data are fully consistent with the decomposition
of this feature proposed in Paper I.
The absorption component is best fitted 
by a damped profile 
with neutral hydrogen column density
$N$(H~I)\,=\,($7.0 \pm 1.5$)$\times 10^{20}$\,cm$^{-2}$ 
centered at $v = -265$\kms, the velocity
where metal lines from species which are dominant ion
stages in H~I regions also have the largest optical depth.
Superposed on the red wing of the damped absorption 
is an emission line with a characteristic shape
(see bottom right-hand panel of Figure 3).
The emission is strongest near $v \simeq +300$\kms,
is cut off abruptly on the blue side, and falls off
more gradually at longer wavelengths.
This composite absorption-emission profile
of \lya\ conforms to the predictions for a line
resonantly scattered in an expanding nebula
(e.g. Tenorio-Tagle et al. 1999).
The \lya\ photons we see are those back-scattered
from outflowing material which lies 
{\it behind} the stars.
Thus the emission mirrors the velocity structure
seen in absorption in front of the stars;
the agreement with the kinematics
of the absorption lines described in \S3.3
is very good. The tail of \lya\ emission
extends to $v \sim +1000$\kms, beyond the maximum velocity 
of $-750$\kms\ seen
in absorption; presumably this extreme velocity gas,
if it is also present in the approaching part of the outflow,
has too low an optical depth
in the metal lines covered by our spectrum to be detected in absorption.

\subsection{Metal Lines}

We derived values of column density $N$ for
ions of interest using two approaches.
The first is the apparent optical depth method
which is worth considering here, despite our
relatively coarse resolution of 58\,km~s$^{-1}$ FWHM,
because the profiles of the 
absorption lines appear to be fully resolved.
The apparent column density of an ion 
in each velocity bin, $N_a(v)$ in units of cm$^{-2}$~(km~s$^{-1}$)$^{-1}$, 
is related to the apparent optical depth in that bin $\tau_a(v)$
by the expression
\begin{equation}
N_a(v) ~=~ \frac{\tau_a(v)}{f\lambda} \times \frac{m_e c}{\pi e^2} 
~=~  \frac{\tau_a(v)}{f\lambda{\rm (\AA)}} \times 3.768 \times 10^{14}
\end{equation}
where the symbols $f$, $\lambda$, $c$, $e$ and $m_e$ have their usual
meanings.
The apparent optical depth is deduced directly 
from the observed intensity
in the line at velocity $v$, $I_{\rm obs}(v)$, by
\begin{equation}
\tau_a(v) ~=~ -ln~ [I_{\rm obs}(v)/I_0(v)]
\end{equation}
where $I_0(v)$ is the intensity in the continuum.
With the assumption of negligible smearing of the 
intrinsic line profile by the 
instrumental broadening function,
\begin{equation}
\tau_a(v) \approx \tau(v)
\end{equation}
and the total column density of ion $X$, $N_{\rm aod}(X)$,
is obtained by direct summation of eq. (2)
over the velocity interval where line absorption
takes place.

As emphasized by Savage \& Sembach (1991),
the attraction of the apparent optical depth method
lies in the fact that no assumption has to be made
concerning the velocity distribution of the absorbers.
Furthermore, it provides a stringent consistency check when 
two or more transitions arising from the same atomic energy level,
but with different values of the product $f \lambda$, 
are analyzed.
The run of $N_a(v)$ with $v$ should be the same, within the errors
in $I_{\rm obs}(v)$, for all such lines.
In general this condition will {\it not} be satisfied
if there are unrecognized saturated components to the absorption lines;
such a situation will manifest itself with the deduced value of 
$N_a(v)$ being smaller for lines with larger values of $f\lambda$.
In our case a similar effect can also result from geometrical
effects due to the fact that the background source
against which the absorption takes place is not point-like,
but is a spatially extended ensemble of O and B stars.
If the covering of the integrated stellar continuum 
by the absorbing gas is inhomogeneous at a given velocity, 
the apparent optical depth method will give discordant
values of column density for different 
transitions of the same ion at that velocity.

Values of $N_{\rm aod}$ deduced for absorption lines which are not 
saturated\footnote{Once $I_{\rm obs}(v)$ approaches 
zero, $\tau_a(v)$ becomes undetermined (see eq. 3)
and the method is no longer applicable.} are listed
in column (5) of Table 2, while column (4) gives the 
velocity interval $\Delta v$ over which 
equation (2) was summed. 
In Figures 4, 5, and 6 we have reproduced
some examples of the run of $N_a(v)$ with $v$
for lines of interest. 
Figure 4 shows the apparent optical depth
profiles of the two ions for which we have the 
largest number of well-observed transitions, Fe~II and Si~II.
The Fe~II lines shown have values of $f \lambda$
spanning a range of a factor of 10, from the weakest
$\lambda 2374$ to the strongest $\lambda 2382$,
while from Si~II~$\lambda 1808$ to Si~II~$\lambda 1260$
the range in $f \lambda$ is a factor of 320.
It can be seen from these plots that the absorption
in the wings of the lines does not satisfy
the consistency check discussed above, in that
the value of $N$($v$) decreases with increasing
$f \lambda$. As explained above, this could be an indication
of either saturated absorption components which
are too narrow to be resolved,
or of inhomogeneous coverage of the stellar continuum. 
The situation is less clear-cut in the main
component centered at $v = -265$\kms.
Here there is good agreement between 
Fe~II~$\lambda 2374$ and $\lambda 1608$
but these two lines differ by only a factor
of 1.25 in their values of $f \lambda$.
S~II~$\lambda\lambda 1250, 1253$ (Figure 5),
with values of $f \lambda$ differing by a factor of 2, 
are also mutually consistent, although the weaker
line is rather noisy.

The most suggestive evidence of inhomogeneous covering
is provided by the
Si~IV doublet lines. As can be seen from the lower panel
of Figure 6, the stronger member of the doublet
consistently underestimates the column density 
at all velocities. The smoothness of the line profiles
favors geometrical effects in the covering factor, 
rather than the presence of a multitude of regularly spaced, narrow, 
unresolved components, as the more likely interpretation.
This would suggest that the {\it spatial}
distribution of the ionized gas is significantly different
from that of the neutral component (where
the first ions of the heavy elements reside),
even though both H~II and H~I gas occur
over the same velocity interval.
Al~III appears to be an intermediate case
(top panel of Figure 6). 

Our second approach to the derivation of column
densities consisted of fitting theoretical
line profiles to only the central component of the
absorption lines;
this component accounts for most of the 
neutral gas to which our subsequent abundance analysis
refers and, as argued above, geometrical
covering factors may have only a secondary effect
in the derivation of the column densities.
We used the software package
Xvoigt (Mar \& Bailey 1995) to compute 
theoretical Voigt profiles, convolve them
with the instrumental resolution,  and compare
them to the data. 
We found that all the Si~II, S~II, and Fe~II lines
could be fitted consistently (that is, 
with the same column density for all transitions
of the same ion, irrespective of the value of
$f \lambda$) with a single absorption component
with velocity dispersion parameter $b = 70$\kms\
centered at $v = -265$\kms. Furthermore, 
C~I~$\lambda 1157$, the 
Mg~II~$\lambda\lambda 1239, 1240$  doublet, P~II~$\lambda 1152$,
Mn~II~$\lambda 2576$ (the strongest member of the triplet and the only one 
detected), and the numerous Ni~II transitions are all consistent
with these parameters, although they do not constrain them
stringently because these weak features are generally noisy and in many
cases blended (see Figure 7). 
The N~I$~\lambda 1134$ and $\lambda 1199$
triplets, which are intrinsically blended because the
three lines are closer in wavelength than the velocity
extent of the absorption,
were also fitted with these values of $b$ and $v$,
and are discussed in more detail in \S5.3 below.

Values of column density $N_{\rm pf}$ deduced by fitting the
profile of the central component are listed in 
column (6) of Table 2.
Comparison with the values of 
$N_{\rm aod}$ in column (5) 
shows that the former are consistently 
below the latter by 0.2 dex. That is,
the central component accounts for 63\%
of the total column density observed. One exception
is C~I for which the two values
of $N$ are in good agreement; inspection of Figure 1
shows that this is because this line extends
over a narrower velocity range than even the weakest lines
from the first ions, and all the absorption
is accounted for by a single $b = 70$\kms\ absorption component
at $v = -265$\kms. \\

\section{ELEMENT ABUNDANCES}
\subsection{Which Column Densities?}

We can summarize our column density determinations as follows.
For all the species considered (except N~I and Mg~II 
which are discussed separately below), we cover at least one transition
which is sufficiently weak for the apparent optical depth method
to be applicable. The column densities derived this way have
a typical statistical error of about $\pm 0.1$\,dex, but are
systematically {\it underestimated} by an unknown amount 
because we see indications in the wings of the line profiles
of either unresolved saturated components or inhomogeneous
coverage of the integrated stellar UV light.
Approximately two thirds of the column density is due to a 
main absorption component at $v = -265$\kms\ which apparently 
lies well in front of the stars and covers them completely,
since the strongest absorption lines have black cores 
near this velocity. 

We do not know what fraction of the total column density of 
neutral hydrogen resides in this component, as our
estimate of $N$(H~I)\,$= 7.0 \times 10^{20}$\,cm$^{-2}$ was derived
by fitting the damping wings of the \lya\ line.
However, this fraction is likely to be less than 100\%
because there is O~I  absorption at velocities beyond 
those attributable to the main component (see bottom right-hand
panel of Figure 1).\footnote{O~I and H~I are closely tied together,
so that $N$(O~I)/$N$(H~I)\,$\simeq$\,(O/H) irrespective of the 
ionized fraction of the gas (Sembach et al. 2000).}
Consequently, in deducing element abundances
we have referred the {\it total} ion column densities to the 
above value of $N$(H~I); the systematic error introduced 
by this assumption amounts to at most $+0.2$ dex (the
ratio between the total column density of ion $X$ and that
due only to the component at $v = -265$\kms).

Our estimate of $N$(H~I) does not include any contribution
from molecular hydrogen. Since our ESI spectrum does not extend 
to the far-UV wavelengths where the strongest H$_2$ transitions
occur, we cannot place empirical limits on the 
molecular fraction of the gas. 
Qualitatively, it is likely that
some H$_2$ is present in the main absorption component,
where we see C~I, because the same photons which would dissociate
the H$_2$ molecules through absorption in the Lyman and Werner bands
would also ionize C~I (Hollenbach, Werner, \& Salpeter 1971;
Federman, Glassgold, \& Kwan 1979).
On the other hand, in local interstellar clouds
with neutral hydrogen column densities
comparable to that in cB58, the molecular fraction 
is seldom greater than 50\%. For example,
Savage et al. (1977)
found $N$(H~I)/[$N$(H~I)+2$N$(H$_2$)]\,$\geq 0.5$
in all but two of the 109 stars they surveyed
with the {\it Copernicus} satellite.
A recent FUSE survey of 70 stars in the 
Small and Large Magellanic Clouds 
(Tumlinson et al. 2002) has shown that 
in the interstellar medium of those galaxies the
molecular fraction is even lower.
Thus, it would be surprising if
the element abundances we deduce below were 
overestimates by more than a factor of two because
of our inability to account for H$_2$,
although obviously this uncertainty can only
be resolved by extending
the spectral study of cB58 to shorter wavelengths.

The values of $N$(X) adopted are listed in column (2) 
of Table 3; these are the same as the values 
of $N_{\rm aod}$ from Table 2 (with appropriate averaging
when more than one line is available for the same ion)
with the exception of P~II, N~I, and Mg~II.
For P~II we cover only one transition, at $\lambda 1152.818$,
which is blended with \lya\ forest absorption within the 
velocity range over which the optical depth is 
integrated; excluding this contamination we deduce
log~$N_{\rm aod} > 14.04$ (cm$^{-2}$).
Profile fitting gives log~$N_{\rm pf} = 14.00$ (cm$^{-2}$)
for the main component only (see Figure 7); 
thus in Table 3 we have adopted 
log~$N$(P~II)\,$= 14.20$ (cm$^{-2}$) to account for 
the remainder of the absorption, as explained above. 
The situation is more complicated
for N~I and Mg~II because the blending here is between lines
of the same multiplet.
When fitting the profiles of these blended features
the wings from adjacent lines within the multiplet
will have contributed somewhat to the absorption
in the main component, and so it is unclear that the values
of $N_{\rm pf}$ should in fact be corrected upwards
by +0.2\,dex in these cases too. We have not done so
and adopted the values of $N_{\rm pf}$ in our
abundance analysis; again the systematic error introduced
is at most a factor of $-0.2$\,dex and does not
affect the main conclusions of our abundance analysis below.

\subsection{Ionization Corrections}

The species listed in Table 3 are the main 
ionization stages of their respective elements
in H~I regions; by taking the ratios of their
column densities relative to $N$(H~I) to give
the abundances listed in column (3) 
we make the implicit assumption that the fractions of
the first ions (and of N~I) which reside in ionized
gas are negligible for our purposes.
At first sight there may seem to be little justification
for this assumption---after all
we know from our data that there is 
ionized gas at the same velocities as
the neutrals and first ions (Figures 1 and 2).

The problem of correcting 
interstellar abundance measurements
for H~II region `contamination'
is not a new one (e.g. Steigman, Strittmatter, \& Williams 1975);
most recently it has been scrutinized in detail
by Howk \& Sembach (1999) and Vladilo et al. (2001)
using the CLOUDY code (Ferland et al. 1998).
Vladilo et al. have specifically calculated the 
relevant corrections as a function of neutral hydrogen 
column density, prompted by the realization that
in damped \lya\ systems (DLAs)
highly ionized species---particularly Al~III---are 
often coincident in velocity with the first ions,
as is the case here. 
These authors confirm the main conclusion 
of previous work, that the corrections to interstellar
abundances due to the presence of ionized gas along the line of sight
are generally small. For the elements considered here,
when log~$N$(H~I)$ = 20.85$ (cm$^{-2}$) as in cB58,
the largest correction is for P~II with
log~(P/H) = log~[$N$(P~II)/$N$(H~I)~$] - 0.1$\,dex.

What the CLOUDY calculations show is that 
when ionizing radiation impinges on the surface
of a neutral interstellar cloud and ionizes its outer layer,
this H~II region contains relatively small fractions
of the first ions compared with the interior, self-shielded,
H~I region (at least when 
$N$(H~I)\,$ \simgt 2 \times 10^{20}$\,cm$^{-2}$), because
higher ionization stages than the first ions dominate.
The situation under consideration here is 
more complex than that considered by Vladilo et al. (2001)
whose calculations did
not include processes such as X-ray
ionization from the very high temperature gas which presumably 
drives the outward motion of the ISM in cB58, nor the 
conduction between the expanding bubble of hot gas 
and the swept-up interstellar matter. Even so,
it is unlikely that these additional sources of ionization
would affect significantly the relative abundances of the 
neutrals and first ions.
On the other hand, ionization corrections 
may be important if the column density 
$N$(H~I)$\,= 7.0 \times 10^{20}$~cm$^{-2}$
we measure in cB58 is due to the superposition
of many individual components,
each with $N$(H~I)$\, < 1 \times 10^{20}$~cm$^{-2}$.

\subsection{Chemical Abundances in the Interstellar Medium of cB58}

In summary, we have examined various sources of systematic uncertainty
which could affect the abundance determinations in Table 3 
and concluded that in general these are likely to be less than
0.2~dex. We have no way of quantifying the effect of saturation
in the wings of the line profiles which may have caused us to
underestimate the overall level of metallicity;
in order to do so we have to rely on comparing
our results with what is already known about the 
degree of metal enrichment of cB58. The last column in Table 3
gives the abundances of the elements observed relative to
the solar reference scale; these are also shown graphically
in the top panel of Figure 8. We now discuss these results
individually.

Our observations cover three $\alpha$-capture elements,
Mg, Si, and S; all three give a consistent picture
with abundances approximately 2/5 solar or
[Mg,Si,S/H]$_{\rm cB58} \simeq -0.4$ in the usual notation.
This is in excellent agreement with the value
[O/H]$_{\rm cB58} \simeq -0.35$ for the {\it ionized}
gas, obtained by Teplitz et al. (2000)
from infrared observations of nebular emission 
lines.\footnote{Using the ratio of [O~II] and [O~III] to H$\beta$
(the $R_{23}$ method of Pagel et al. 1979), 
Teplitz et al. found log~(O/H)\,$ \simeq -3.61$.
With the recent revision of the solar oxygen abundance
proposed by Holweger (2001), log~(O/H)$_{\odot} = -3.26$,
this leads to [O/H]$_{\rm cB58} \simeq -0.35$\,.}
The agreement, unless entirely fortuitous,
gives confidence in our abundance analysis and shows
that the systematic uncertainties 
discussed at length above have not been underestimated in our 
treatment. The abundance of P, [P/H]$_{\rm cB58} \simeq -0.21$, 
is in line with those of Mg, Si, and S, given the errors.

Nitrogen, on the other hand, is apparently ten times 
less abundant than the $\alpha$ elements relative to the solar scale,
and $\sim 3$ times below the typical log~(N/O)\,$ = -1.35$
measured in nearby galaxies when [O/H]\,$ = -0.35$
(Henry, Edmunds, \& K\"{o}ppen 2000).\footnote{For comparison,
log~(N/O)$_{\odot} = -0.81$ (Holweger 2001). The dependence
of (N/O) on [O/H] results from the complex nucleosynthesis
of N, which has both a primary and a secondary component---see
the discussion by Henry et al. (2000).}
The low abundance of N we derive 
contrasts with the results of the nebular analysis 
by Teplitz et al. (2000). These authors used their
measurement of the [N~II]/[O~II] ratio to deduce
log~(N/O)\,$ = -1.24$ or [N/H]$_{\rm cB58} \simeq -0.78$,
whereas we find [N/H]$_{\rm cB58} \simeq -1.43$, a difference
of a factor of 4.5\,.
To explore the origin of this discrepancy we used Xvoigt
to compute the absorption profiles of the N~I~$\lambda 1134$ 
(see Figure 9) and $\lambda 1199$ triplets 
produced by the main absorption component of the 
interstellar lines at $v = -265$\,\kms\ with $b = 70$\,\kms\ 
and $N$(N~I)\,$ = 1.0 \times 10^{16}$\,cm$^{-2}$, the column density
of N~I corresponding to the value [N/H]$_{\rm cB58} \simeq -0.78$
derived by Teplitz et al. (2000).
As can be seen from Figure 9, such a high value
of N~I is highly inconsistent with  
our absorption line data. 

We have considered a number of possible explanations 
for the difference in the N abundance deduced from 
absorption and emission line measurements.
First, the conversion from the 
emission line ratios to (N/O) has only a minor
temperature dependence 
which cannot account for the factor of 4.5 discrepancy.
Second, the possibility that there is a real composition
difference between ionized and neutral gas seems
unlikely because (a) there are no precedents for such
abundance inhomogeneities in local galaxies
(Meyer, Cardelli, \& Sofia 1997; Esteban et al. 1998; Kobulnicky 1999),
and (b) the abundances of the $\alpha$ elements
Mg, Si, and S we derive for the neutral component
are all in excellent agreement with that of O in the H~II gas.
Third, one may conjecture that N may be more highly ionized than H,
if the H~I region is permeated by extreme ultraviolet photons
produced from stars and hot gas, since the 
cross section for photoionization of N~I to N~II is higher than
that of H~I. Such ionization effects have been considered in detail
by Sofia \& Jenkins (1998) and Jenkins et al. (2000) who found them
to be important only when $N$(H~I)\,$\simlt 10^{19}$\,cm$^{-2}$,
that is at column densities of neutral hydrogen $\sim 100$ times lower
than that seen in cB58. When $N$(H~I)\,$ > 10^{20}$\,cm$^{-2}$,
the interstellar ratio N~I/H~I is in good agreement
with the solar abundance of N (Sofia \& Meyer 2001a,b).
In the nearby starburst galaxy NGC~1705, 
where $N$(H~I)\,$\simeq 1.5 \times 10^{20}$\,cm$^{-2}$,
Heckman et al. (2001b)
were able to address this point directly with observations of both
N~I and N~II, and indeed they found that N~I/N~II\,$\simeq$\,H~I/H~II\,.
We conclude that ionization effects are unlikely to 
have led us to underestimate significantly the abundance of N
in our absorption line analysis.

On the other hand, examining the NIRSPEC spectra published by Teplitz et al.
(see their Figure 3), it can be realized
that the weak [N~II]~$\lambda\lambda 6548, 6583$ lines
are right at the edge of their data and,
at observed wavelengths of 2.44204 and 2.45566 $\mu$m respectively, 
well beyond the conventional boundary of the $K$-band window, in a region
where both atmospheric absorption and thermal emission
are high. It is conceivable, therefore, that their 
fluxes may have been overestimated; the problem is compounded
by the awkward location of [O~II]~$\lambda 3727$ ($\lambda_{\rm obs} = 1.3898\, \mu$m)
between the $J$ and $H$ windows, close to the maximum of H$_2$O absorption
at these wavelengths. It is also true that among our abundance determinations 
that of N is the least precise, because the individual lines in the two triplets
are not resolved, but any additional absorption components which have been omitted
in the profile fit shown in Figure 9
would probably have the net effect of reducing, rather
than increasing the value of $N$(N~I).

Turning to the Fe-peak group, we have 
abundance measurements for three elements,
Mn, Fe, and Ni. As can be seen from Figure 8, they are all less
abundant than S and the other $\alpha$-capture elements, by factors of between 
0.4 (Ni) and 0.75 (Fe) dex. This could be a real departure from
the solar scale, or a reflection of the common depletions
of Mn, Fe, and Ni onto dust grains, or both.
The ambiguity could in principle be resolved by consideration of the
abundance of Zn, an iron-peak element which is normally
undepleted (e.g. Pettini, Boksenberg, \& Hunstead 1990).
Unfortunately, we cannot derive a useful measure of 
[Zn/H] because the Zn~II doublet lines are 
blended with atmospheric $B$-band absorption and with
stellar photospheric lines.

We can still explore this question in a less direct way by
comparing the pattern of element abundances in cB58
with that of the interstellar gas near the Sun
as reviewed by Sembach \& Savage (1996) and Savage \& Sembach (1996).
These authors give illustrative cases of the depletions 
of refractory elements observed in different
interstellar environments, from diffuse, cool clouds in the disk
(where the depletions are most pronounced) to
low-density halo sight-lines (where depletions are less
severe). In Figure 8 we compare the pattern of abundances 
in cB58 (top panel) with that typically found along
Galactic sight-lines which intercept warm clouds 
in the disk and halo (lower panel).\footnote{We have chosen 
this example simply because the deficiency of Fe is similar
to that found in cB58 and data are available for several
elements. Our conclusion would still be the same if we compared 
with one of the other Galactic depletion patterns which
roughly match the Fe abundance in cB58.}
To facilitate the comparison,
we show the Galactic measures (thin rectangles) shifted to lower values by
the $-0.40$\,dex intrinsic underabundance of the $\alpha$-elements 
in cB58 (thick rectangles).
The relative abundance of Ni
in the lower panel of Figure 8
has been increased by 0.27 dex to account for the downward revision 
of the $f$-values of the relevant transitions 
(Fedchak, Wiese, \& Lawler 2000 and references therein)
since the compilation by Savage \& Sembach (1996).
The abundances of N and P, not included in the compilation
by Savage \& Sembach, are from the surveys
by Meyer et al. (1997) and Dufton, Keenan, \& Hibbert (1986) respectively.

From the lower panel of Figure 8 it can be realised
that, if depletion were the sole origin 
of the observed underabundances 
of the Fe-peak elements in cB58 (relative to the undepleted S),
then we would also expect Mg, Si, and P to be depleted, albeit to
a lesser degree. This is not what is observed in cB58
(top panel), where all three elements are 
approximately solar relative to S.
Of course there is no reason why the composition of dust grains in 
the compressed, outflowing, ISM of cB58 should be the same as that
seen in nearby Galactic clouds. Furthermore, some dust {\it must}
be present in the ISM of cB58, since the UV continuum is significantly
reddened (Paper I) and dust emission has been detected directly at
sub-mm and mm wavelengths (van der Werf et al. 2000; Baker et al. 2001).
On the other hand, 
the fact that all the observed Fe-peak elements are uniformly 
below the $\alpha$-capture group---coupled
with the obviously instrinsic (i.e. not due to dust depletion)
underabundance of N---is very suggestive of 
more fundamental differences
between this high redshift galaxy and the Milky Way
in their histories of element production, as we now discuss.
The most likely scenario, in our view, is that both dust depletion
and intrinsic departures from the solar scale contribute 
to the pattern of element abundances we have uncovered in cB58.

\section{DISCUSSION}

\subsection{The Star Formation History of cB58}

What do these results tell us about the history of star formation
in cB58? The most immediate and straightforward conclusion
is that this galaxy had already achieved a near-solar metallicity
at relatively early times, some 12\,Gyr ago in 
a $\Omega_{M}=0.3$, $\Omega_{\Lambda} = 0.7$, $h = 0.65$
cosmology, when the age of the universe was only 17\%
of what it is today.
The results of our quantitative analysis of the interstellar absorption
lines confirm earlier inferences 
of a high metallicity based on the strengths 
of the stellar wind lines, which exhibit
P Cygni profiles of comparable extent to those seen
in O stars of the Large Magellanic Cloud (Paper I; Leitherer et al. 2001).
The advanced degree of chemical enrichment we have established
is consistent with the original
suggestion by Steidel et al. (1996) that, in the Lyman
break galaxies, we see the progenitors of
today's ellipticals and bulges, and with a general
prediction of hydrodynamic cosmological simulations
(e.g. Cen \& Ostriker 1999; Nagamine et al. 2001)
which show that such relatively high abundances are in fact common
in the most massive galaxies at $z \simeq 3$.
Of course we do not know whether in this respect cB58 is typical
of the Lyman break galaxy population in general.
However, its nebular emission line ratios {\it are}
very much in line with those of other 
$L^{\ast}$ LBGs studied by Pettini et al. (2001).
In that paper we showed how the $R_{23}$ values
measured admit a range of values of 
[O/H], because of the double-valued nature of this
abundance calibrator. If cB58 is a typical member of the sample,
solutions on the upper branch of the $R_{23}$ vs. [O/H]
relation are favored, corresponding to near-solar
abundances of oxygen.

The second conclusion from the present study is that
elements whose release into the ISM is delayed relative to 
the $\alpha$-capture products of Type II supernovae 
are significantly underabundant in cB58. The most obvious example
is N for which dust depletion is not an issue, but
we have argued that even the refractory Fe-peak
elements probably exhibit a degree of intrinsic
deficiency relative to the solar pattern of abundances.
In standard chemical evolution models, intermediate mass stars
are the main source of N, while Type Ia supernovae 
contribute most of the Fe-peak elements. In both cases
the evolutionary times are significantly longer than 
the near-instantaneous cycling by massive stars which explode
as Type II supernovae, with the consequence that the 
mixing of N and Fe-group elements into the ISM is delayed
by between 300\,Myr and 1\,Gyr.\footnote{An alternative way to
reduce the relative abundances of elements synthesized by
intermediate mass stars is to appeal to a top heavy, or truncated,
initial mass function (IMF). Although
we cannot exclude this possibility, we consider it less
likely, given the lack of positive
evidence to date against the simplest assumption of
a universal IMF (Kennicutt 1998; Feltzing, Gilmore, \& Wyse 1999;
Molaro et al. 2001).}

The results of our analysis would then suggest a relatively 
young age for cB58. Ellingson et al. (1996) reached a similar conclusion
by fitting the rest-frame UV-optical spectral energy distribution (SED)
of the galaxy with spectral synthesis models, favouring an age 
of less than 35\,Myr.
However, this only applies to the 
stellar population responsible for most of the UV-optical
light we see; longer wavelength photometry would be required
to recognize an older population
which has now `switched-off', even if it accounted
for as much as 90\% of the galaxy mass. This is a common
limitation in the diagnostic value of SEDs when only a relatively
narrow range of wavelengths is available 
(e.g. Papovich, Dickinson, \& Ferguson 2001).
The abundance data presented here, on the other hand, when
interpreted in the context of current ideas of stellar nucleosynthesis,
rule out the existence of a significant population of stars older
than $\sim 300$\,Myr---had such an earlier episode of 
star formation taken place,
it would have enriched the ISM
of cB58 in N above the observed level.
Unpublished {\it ISOCAM} photometry,
covering the rest frame near-IR region
from 1 to 4\,$\mu$m, is consistent with this picture;
in their preliminary conference report Bechtold et al. (1998)
place an upper limit of $\sim 10$\% to the fraction
of the stellar mass contributed by a 2 Gyr-old population.

The spectral character of 
cB58---with its reddened UV continuum, 
strong saturated interstellar lines, and very 
weak \lya\ emission---fits in well with the scenario proposed
by Shapley et al. (2001) where these are precisely the 
youngest galaxies in the Lyman break population.
Thus, {\it in cB58 we may well have an empirical example
of a galaxy caught in the act of converting
a substantial fraction of its interstellar gas
into stars on a few dynamical timescales.}
Incidentally,
our value of [N/$\alpha$] resolves the apparent
discrepancy noted by Teplitz et al. (2000)
between the relatively high (N/O) they derived and
the young age deduced by Ellingson et al. (1996).
Precise abundance determinations
from weak nebular emission lines are difficult once these features
are redshifted into the near-IR, as is the case in LBGs.
Such work will probably have to await the 
availability of the Next Generation Space Telescope
to develop fully.

\subsection{Comparison with Damped \lya\ Systems}

The pattern of element abundances 
in Figure 8 is strikingly different from that typical of 
damped \lya\ systems. First, DLAs at all redshifts are generally
metal poor and the chance of finding one with 
metallicity $Z_{\rm DLA} \simeq 2/5 Z_{\odot}$
at redshifts $z = 2-3$ is only about one in ten
(Pettini et al. 1999). Second, and somewhat paradoxically, the enhancement
of the $\alpha$-elements relative to Fe {\it expected}
at these low metallicities by analogy with metal-poor stars
in the Milky Way has proved very elusive to pin-down
in DLAs (Vladilo 1998; Pettini et al. 2000b; Prochaska \& Wolfe 2002).
The data for N are less extensive, but again the majority
of DLAs seem to have (N/O) ratios in line with those 
of today's metal-poor galaxies where there has been sufficient time
for the dispersal of the primary N from stars of intermediate mass
(Henry et al. 2000; Lattanzio et al., in preparation).

There are also obvious differences in the kinematics
of the absorbing gas between DLAs and cB58. 
Metal lines seldom span more than 
300\kms\ in damped systems (Prochaska et al. 2001), 
whereas we have found that in cB58 
the same transitions extend over an interval
$\Delta v \simeq 1000$\kms. Furthermore,
while all ion stages observed in cB58
broadly share the same velocity structure,
this is not generally the case in DLAs.
Here it is common to find significant differences
in the kinematics of high and low ions;
Wolfe \& Prochaska (2000) have taken this as evidence for
distinct sub-structures, such 
as infalling ionized clouds and 
a neutral thick disk, within the same dark matter halo.

These chemical and kinematic differences can be understood
if star formation proceeds at a much faster pace in LBGs compared
with the DLAs, and indeed several recent theoretical studies have proposed
such a picture (e.g. Mo, Mao, \& White 1999; Jimenez, Bowen, \& Matteucci 1999).
What is surprising, perhaps, is that there should be so little
overlap between the properties of these two populations (as far as we can tell on the
basis of the available, limited, set of data), almost as if their
modes of star formation were qualitatively different.

\subsection{Outflows in Lyman Break Galaxies}

As we have seen, the bulk of the ISM of cB58, rich
in the accumulated products of stellar nucleosynthesis, 
is moving outwards from the star forming region with a velocity 
$v \simeq 255$\kms; furthermore, there is gas with outflow speeds 
up to $3-4$ times higher seen both in interstellar absorption
and \lya\ emission.
In Paper I we showed that, if the column density $N$ is 
confined to a spherically symmetric thin shell of radius $d$ 
expanding with velocity $v$,
the implied mass outflow rate is
\begin{equation}
\dot{M}  \sim 70 \times 
\left( \frac{d}{1~{\rm kpc}} \right)  \times 
\left( \frac{N}{7 \times 10^{20}~{\rm cm}^{-2}} \right)  \times
\left( \frac{v}{255~{\rm km~s}^{-1}} \right)~
M_{\odot}~{\rm yr}^{-1}.
\end{equation}
We also pointed out that this value is
comparable to the star formation rate
SFR\,$\simeq 40 M_{\odot}$~yr$^{-1}$ deduced from
the far-UV luminosity $L_{1500}$ after correcting
for a factor of $\sim 7$ attenuation by dust extinction.
While we have no direct information on the dimensions
of the shell, the fact that the absorbing material 
seems to cover completely the stars
suggests that $d$ may well be $>1$\,kpc.
As mentioned in \S3.3, the lensing analysis
by Seitz et al. (1998) favoured a source size
$r_{1/2} = 2 h_{70}^{-1}$\,kpc, where $r_{1/2}$ is the half-light radius
in the rest-frame UV;
if $d \simgt 2r$, $\dot{M}$ in equation (5) 
may well be many times greater than the SFR,
unless we are seeing a highly collimated outflow.
This conclusion would be further strengthened 
by including in the calculation the higher velocity gas,
if its ionized fraction is high so that
$N$(H~II)\,$ > N$(H~I) at $v \gg 255$\,\kms.

Such powerful superwinds are seemingly common in
LBGs---the 255\kms\ blueshift of the interstellar
lines in cB58 is certainly typical of the sample of LBGs
with luminosities $L \geq L^{\ast}$
(Pettini et al. 2001; Adelberger et al., in preparation).
Comparable outflows are seen in local starburst
galaxies when the star formation rate per unit 
area exceeds $0.1 M_{\odot}$~yr$^{-1}$~kpc$^{-2}$
(Heckman et al. 2000). At both high and low redshifts
the impact of these outflows on their host galaxies
and on the surrounding environment must be 
considerable.
Quantitatively, an important question is whether
the superwinds and the material entrained in the flow
escape the galaxies altogether
or remained confined within their gravitational
potential wells. Aguirre et al. (2001) have,
among others, addressed this question 
with the aid of hydrodynamical simulations
and concluded that the most important
factors determining the stall radius of the 
shell are the wind velocity, the mass of the
wind-driving galaxy, and the fraction 
of ambient ISM entrained.

We can roughly estimate the mass of cB58 as follows.
Under the simplest assumption that the star formation rate 
has remained constant for a period between 
$\sim 30$ and $\sim 300$\,Myr,
the formed stellar mass is
\begin{equation}
m_{\rm star} \simeq 4 \times 10^9 \, t_{\rm sf8}\, M_{\odot}
\end{equation}
where $t_{\rm sf8}$ is the duration of the star formation
activity in units of 100\,Myr.
In order to reach a metallicity $Z \simeq 2/5 Z_{\odot}$,
at least 1/3 of the gas reservoir must have been processed 
into stars (for a solar yield; Edmunds 1990). Thus,  
\begin{equation}
m_{\rm baryons} \simlt 1.2 \times 10^{10} \, t_{\rm sf8}\, M_{\odot}
\end{equation}
This is comparable to the dynamical mass
\begin{equation}
m_{\rm dyn} \simeq 1.3 \times 10^{10}\,M_{\odot}
\end{equation}
for the starburst region
obtained by substituting into the expression
\begin{equation}
m_{\rm dyn} = 1.2 \times 10^{10} M_{\odot} \left( \frac{\sigma}{100\,{\rm km~s^{-1}}} \right)^2 \frac{r_{1/2}}{{\rm kpc}}
\end{equation}
(Pettini et al. 2001)
the velocity dispersion of the nebular emission lines
$\sigma = 81$\kms\ measured by Teplitz et al. (2000)
and $r_{1/2} = 2 h_{70}^{-1}$
deduced by Seitz et al. (1998).

The simulations by Aguirre et al. (2001) predict that
for masses of order $10^{10}\,M_{\odot}$ and 
{\it initial} outflow speeds 
of $\sim 300$\kms\ the stall radius 
is relatively small, $r_{\rm stall} \simlt 10$\,kpc.
However, if the absorbing gas we see moving
at $v = -255$\,\kms\ is located several kpc
in front of the stars, as we have argued above, 
then (a) its initial velocity must have been significantly
higher, and (b) it has already climbed out of much of the
potential well. For example, the escape velocity
from $d = 4$\,kpc within a Navarro, Frenk, \& White (1997)
halo of mass $10^{10}$ ($10^{11}$) $M_{\odot}$ is 
$\sim 140$ ($\sim 350$) \kms\ (Adelberger et al., in preparation), 
well within the velocities we measure.
Empirically, we can only say that cB58 must have been
able to retain at least some---if not most---of 
the heavy elements synthesized by earlier star formation episodes,
given that we now see them mixed in the ambient, neutral ISM.

More generally, however, the work by Adelberger et al. (in preparation)
show that outflows from LBGs have a detectable impact 
on the intergalactic medium (IGM) out to large radii, of order 100\,kpc. 
Together with the wide range of ages of LBGs deduced by Shapley et al. (2001),
these results may help explain the presence of metals in the \lya\ forest
up to highest redshifts probed ($z = 5$; Songaila 2001).
This raises an interesting possibility.
If the rapid processing of gas into stars we have found in cB58
were common to most LBGs, it is possible that, in at least some cases, 
the gas polluting the IGM has a distinctly non-solar composition,
with a relative deficiency of the elements synthesized
by intermediate and low mass stars. While there are clues
that N may indeed be underabundant relative to C and Si in the
\lya\ forest at $z > 3$, (Boksenberg \& Sargent, 
in preparation), in general it is difficult
to distinguish abundance anomalies from other factors,
such as the shape of the ionizing radiation field,
which contribute to the relative column densities
of the few ion stages observed in the forest. 

Finally, the impact of violent star formation on the
ambient ISM of cB58 will not be limited to kinematic effects.
It is also likely to produce regions where
the gas is optically thin to Lyman continuum photons
which can thus escape into the IGM and 
contribute to the ionizing background 
(Steidel, Pettini, \& Adelberger 2001).
Heckman et al. (2001a) have recently used the example of cB58
to argue that superwinds do not necessarily
clear out such channels,
by pointing out that if the metal lines are optically thick
in their cores, the optical depth at the Lyman edge
is likely to be orders of magnitude larger.
Indeed, on the basis of the results presented here,
we would not expect to detect any flux below 912\,\AA\ in cB58;
although we argued earlier that the covering factor
of some absorption components may not be unity,
there is little doubt that the bulk of the ISM 
moving at $v = -255$\,\kms\ is opaque to ionizing radiation
{\it as viewed from our line of sight}. 
Nevertheless, this does not preclude the possibility
that the UV light from O and B stars may be escaping
in other directions. 
A local example is provided
by NGC~1705 where the strongest interstellar absorption lines
have black cores, and yet there is evidence that the 
superbubble created by the starburst has begun to break
out of the galaxy and fragment, allowing the hot
interior gas to vent out into the IGM (Heckman et al. 2001b).\\

\section{SUMMARY AND CONCLUSIONS}

We have presented our analysis of the interstellar spectrum of 
MS~1512$-$cB58, an $L^{\ast}$ Lyman break galaxy at 
$z = 2.7276$ which is undergoing active star formation
at a rate of $\sim 40 M_{\odot}$~yr$^{-1}$.
The observations, which have a spectral resolution
of 58\kms, have been made possible by the gravitationally
lensed nature of cB58 and the high efficiency
of the new Echelle Spectrograph and Imager on the Keck~II 
telescope. Our main findings are as follows.

1. The ambient interstellar medium of cB58 is highly
enriched in the chemical elements released by Type~II
supernovae; O, Mg, Si, P, and S all have abundances
of $\sim 2/5$ solar. Thus, even at this relatively
early epoch (corresponding to a look-back time
of 83\% of the age of the universe in today's favored 
cosmology), this galaxy had already processed at least 
one third of its gas into stars.

2. The galaxy appears to be chemically young, in that
it is relatively deficient in elements produced by 
intermediate mass stars whose evolutionary timescales are
longer than those of Type~II supernova progenitors. 
N and the Fe-peak elements
we observe (Mn, Fe, and Ni) are all less abundant
than expected by factors between 0.4 and 0.75\,dex.
Depletion onto dust, which is known to be present in cB58,
probably accounts for some of the Fe-peak element
underabundances, but this is not likely to be an important
effect for N. On the basis of current ideas of the nucleosynthesis 
of N, it would appear that much of the ISM enrichment in cB58
has taken place within the last 300\,Myr, the lifetime
of the intermediate mass stars believed to be the
main source of N. For comparison, the starburst episode
responsible for the UV and optical light we see
is estimated to be younger than $\sim 35$\,Myr,
on the basis of theoretical models of the
spectral energy distribution at these wavelengths.

3. Taken together, these two findings are highly
suggestive of a galaxy caught in the act of converting
its interstellar medium into stars on a few dynamical
timescales---quite possibly in cB58 we are witnessing the 
formation of a galactic bulge or an elliptical galaxy.
The results of our chemical analysis are consistent
with the scenario proposed by Shapley et al. (2001),
whereby galaxies whose UV spectra are dominated by
strong, blueshifted, absorption lines, as is the case 
here, are at the young end of the range of ages of LBGs.
Our findings also lend support to models
of structure formation which predict that, even at $z \simeq 3$,
near-solar metallicities should in fact be common 
in galaxies with masses greater than $\sim10^{10} M_{\odot}$
(e.g. Nagamine et al. 2001).
For cB58 we deduce a baryonic mass 
$m_{\rm baryons} \simeq 1 \times 10^{10} M_{\odot}$,
from consideration of its star formation history,
metallicity, and velocity dispersion of the ionized gas.

4. The chemical properties of cB58 are markedly
different from those of most damped \lya\ systems
at the same redshift. DLAs are significantly
less metal-rich, typically by about one order of magnitude,
and seldom exhibit the enhancement of the
$\alpha$-capture elements we have uncovered here.
Presumably star formation is a slower and more protracted
process in DLAs than in LBGs, again consistent 
with current theoretical ideas about these two
populations of high redshift galaxies.

5. The interstellar medium in cB58 has been stirred to
high velocities by the energetic star formation activity.
In the strongest metal transitions, absorption extends
over a velocity interval of $\sim 1000$\,\kms.
The net effect is an outflow of the bulk of the ISM
at a speed of $\sim 255$\,\kms\ which we see both in absorption 
in front of the stars and in emission of \lya\ photons 
back-scattered from gas behind the stars. There seems to be
a high degree of symmetry in the kinematics of approaching and receding
portions of the flow; this and other empirical properties
we have delineated will provide constraints
to future modeling of such superwinds from starburst galaxies.

6. Unless the outflow is highly collimated, the mass outflow rate
exceeds the star formation rate. 
It is unclear whether the outflowing gas is still bound
to the galaxy. We suspect that its velocity exceeds the
escape velocity within the potential well of cB58 but,
on the other hand, the high metallicity of the ambient ISM 
suggests that most of the elements synthesized by previous
generations of stars have been retained.
In general, outflows from LBGs seem the likely
source of at least some of the metals present 
in the \lya\ forest up
to the highest redshifts probed.
Although the ISM of cB58 is optically thick to Lyman continuum
photons along our line of sight, there remains the possibility that
ionizing photons are leaking out of the galaxy in other
directions, through channels cleared out in the highly
disturbed interstellar medium.

While we have made some progress in measuring
some of the physical parameters
of this Lyman break galaxy,
we remain acutely aware of the fact that we are 
dealing with a single object. 
How typical is cB58 of the Lyman break population
as a whole? There are indications that some of the above
results, particularly the high metallicity, the rapid star formation
timescale, and the kinematics of the outflow, may well be
common to at least a subset of LBGs (Pettini et al. 2001; Shapley et al. 2001).
However, extending the detailed analysis presented here to
other high redshift galaxies, targeting a variety
of spectral morphologies, remains a high priority for 
future work. Obviously there is a strong incentive to
identify further cases of  gravitationally lensed
LBGs, but few sources will be as 
favorably located relative to the lens as 
cB58. Possibly the stellar, rather than interstellar,
spectrum will provide the means to study 
the chemical abundances and other
properties of Lyman break galaxies in a more wholesale manner;
we intend to address this question in a future paper.\\

We are grateful to the ESI team for providing us with
the efficient instrument which made this study possible,
and to the staff of the Keck Observatory for their competent
assistance with the observations. 
We are indebted to Ken Sembach for many suggestions which improved
the paper significantly, and to Andrew Baker, Martin Haehnelt, Tim Heckman,
Francesca Matteucci, and Jeremiah Ostriker
for valuable discussions.
CCS, KLA, MPH, and AES 
have been supported by grants AST95-96229 and AST-0070773 
from the U.S. National Science
Foundation, and by the David and Lucile Packard Foundation.
Finally, we wish to extend special thanks
to those of Hawaiian ancestry on whose sacred mountain 
we are privileged to be guests. Without their generous
hospitality, the observations presented herein would not
have been possible.


%
%

\begin{deluxetable}{lllclrrl}
\tighten
\tablewidth{0pt}
\scriptsize
\tablecaption{\textsc{Interstellar Absorption Lines}\label{tab:isab}}
\tablehead{
  \multicolumn{1}{c}{Ion}
& \multicolumn{1}{c}{$\lambda_{\rm lab}$\tablenotemark{a}}
& \multicolumn{1}{c}{$f$}
& \multicolumn{1}{c}{$z_{\rm abs}$\tablenotemark{b}}
& \multicolumn{1}{c}{$\Delta v$\tablenotemark{c}}
& \multicolumn{1}{c}{$W_0$\tablenotemark{d}}
& \multicolumn{1}{c}{$\sigma$\tablenotemark{d}}
& \multicolumn{1}{c}{Comments} \\
  \colhead{ }
& \multicolumn{1}{c}{(\AA)}
& \colhead{ } 
& \colhead{ } 
& \multicolumn{1}{c}{(km~s$^{-1}$)}
& \multicolumn{1}{c}{(\AA)}
& \multicolumn{1}{c}{(\AA)}
& \colhead{ } 
}

\startdata
C I	& 1157.1857	& 0.5495	& 2.7241   & $-475$ to +\phn25 & 0.18 & 0.05 & \\ 
C II	& 1334.5323	& 0.1278	& 2.7245   & $-775$ to +325 & 3.45 & 0.04 & Blended with C~II*~$\lambda 1335.7$.\\
C IV	& 1548.204	& 0.1908	& 2.7243   & $-775$ to +225 & 4.17 & 0.05 & $W_0$ refers to C~IV~$\lambda 1549$ {\em doublet\/}.\tablenotemark{e} \\
C IV	& 1550.781	& 0.09522	& 2.7245   & $-775$ to +225 & 4.17 & 0.05 & $W_0$ refers to C~IV~$\lambda 1549$ {\em doublet\/}.\tablenotemark{e} \\
N I     & 1134.1653	& 0.01342	& B        & $-475$ to +\phn25 & 1.42 & 0.07 & $W_0$ refers to N~I~$\lambda 1134$ {\em triplet\/}. \\
N I     & 1134.4149	& 0.02683	& B        & $-475$ to +\phn25 & 1.42 & 0.07 & $W_0$ refers to N~I~$\lambda 1134$ {\em triplet\/}. \\
N I     & 1134.9803	& 0.04023	& B        & $-475$ to +\phn25 & 1.42 & 0.07 & $W_0$ refers to N~I~$\lambda 1134$ {\em triplet\/}. \\
N I     & 1199.5496	& 0.1328	& B        & $-475$ to +\phn25 & 2.18 & 0.04 & $W_0$ refers to N~I~$\lambda 1199$ {\em triplet\/}. \\
N I     & 1200.2233	& 0.08849	& B        & $-475$ to +\phn25 & 2.18 & 0.04 & $W_0$ refers to N~I~$\lambda 1199$ {\em triplet\/}. \\
N I     & 1200.7098	& 0.04423	& B        & $-475$ to +\phn25 & 2.18 & 0.04 & $W_0$ refers to N~I~$\lambda 1199$ {\em triplet\/}. \\
N V	& 1238.821	& 0.1570	& 2.7244   & $-425$ to $-125$ & 0.23 & 0.04 & \\	
N V     & 1242.804	& 0.07823	& 2.7242   & $-425$ to $-125$ & 0.23 & 0.03 & \\
O I	& 1302.1685	& 0.04887	& 2.7242   & $-775$ to +125 & 4.25 & 0.04 & $W_0$ refers to O~I~$\lambda 1302$, Si~II~$\lambda 1304$ blend. \\
Mg I    & 2026.4768     & 0.1154	& B        & $-475$ to +\phn25 & 0.57 & 0.02 & $W_0$ refers to Zn~II~$\lambda 2026$, Mg~I~$\lambda 2026$ blend. \\
Mg II	& 1239.9253     & 0.00062	& B        & $-350$ to $-200$  & 0.15 & 0.03 & $W_0$ refers to Mg~II~$\lambda 1240$ {\em doublet\/}. \\ 
Mg II   & 1240.3947     & 0.00035       & B        & $-350$ to $-200$  & 0.15 & 0.03 & $W_0$ refers to Mg~II~$\lambda 1240$ {\em doublet\/}. \\
Al II	& 1670.7886	& 1.833		& 2.7239   & $-775$ to +125 & 2.75 & 0.02 & \\	
Al III	& 1854.71829	& 0.5602	& (2.7244) & $-775$ to +225 &$>$1.90 &$>$0.03 & Partial blend.\\
Al III	& 1862.79113	& 0.2789	& 2.7239   & $-775$ to +225 & 1.54 & 0.03 & \\
Si II	& 1260.4221	& 1.007		& (2.7238) & $-775$ to +125 & 2.80 & 0.04 & Blended with S~II~$\lambda 1259.519$. \\
Si II	& 1304.3702	& 0.094		& 2.7242   & $-775$ to +125 & 4.25 & 0.04 & $W_0$ refers to O~I~$\lambda 1302$, Si~II~$\lambda 1304$ blend. \\
Si II	& 1526.70698	& 0.130		& 2.7239   & $-775$ to +125 & 2.59 & 0.03 & \\
Si II	& 1808.01288	& 0.00218	& 2.7245   & $-475$ to +\phn25 & 0.53 & 0.02 & \\
Si II*	& 1264.7377	& 0.9034	& 2.7234   & $-775$ to +\phn25 & 0.78 & 0.03 & \\
Si II*	& 1309.2757	& 0.094 	& 2.7239   & $-475$ to +\phn25 & 0.15 & 0.03 & \\
Si II*	& 1533.4312	& 0.130		& 2.7232   & $-775$ to +125 & 0.69 & 0.03 & \\
Si III	& 1206.500	& 1.669		& 2.7242   & $-775$ to +225 & 3.23 & 0.06 & \\
Si IV	& 1393.76018	& 0.5140	& 2.7244   & $-775$ to +225 & 2.16 & 0.03 & \\
Si IV	& 1402.77291	& 0.2553	& 2.7242   & $-775$ to +225 & 2.01 & 0.03 & \\
P II    & 1152.818      & 0.2361        & (2.7248) & $-475$ to +\phn25 &$>$0.22 & $>$0.04 & Partial blend (in Ly$\alpha$ forest).\\
S II	& 1250.584	& 0.005453	& (2.7244) & $-475$ to +\phn25 &$>$0.17 &$>$0.03 & Partial blend.\\
S II	& 1253.811	& 0.01088	& 2.7248   & $-475$ to +\phn25 & 0.52 & 0.03 & \\
Cr II	& 2062.2361	& 0.0780	& B        & $-475$ to +\phn25 & 0.35 & 0.03 & $W_0$ refers to Cr~II~$\lambda 2062$, Zn~II~$\lambda 2062$ blend. \\
Mn II   & 2576.877      & 0.3508	& 2.7245   & $-475$ to +\phn25 & 0.38 & 0.10 & \\
Fe II   & 1144.9379     & 0.106         & (2.7239) & $-775$ to +125 & $>$1.20 & $>$0.07 & Partial blend (in Ly$\alpha$ forest).\\
Fe II	& 1608.45085	& 0.058		& 2.7239   & $-775$ to +125 & 1.30 & 0.04 & \\
Fe II	& 2344.2130	& 0.114		& 2.7239   & $-775$ to +125 & 2.99 & 0.04 & \\
Fe II	& 2374.4603	& 0.0313	& (2.7241) & $-775$ to +125 & 1.94 &$>$0.07 & Line affected by bad pixels. \\
Fe II	& 2382.7642	& 0.300		& (2.7239) & $-775$ to +125 & 3.41 &$>$0.05 & Line affected by bad pixels. \\
Ni II	& 1317.217	& 0.07		& 2.7249   & $-475$ to +\phn25 & 0.20 & 0.03 & \\
Ni II	& 1370.132	& 0.0769	& 2.7241   & $-475$ to +\phn25 & 0.24 & 0.02 & Blended with Si~IV~$\lambda 1393$ at $z_{\rm abs}=2.6606$.\\ 
Ni II	& 1703.4119 	& 0.0060	& 2.7244   & $-475$ to +\phn25 & 0.17 & 0.02 & \\
Ni II	& 1709.6042	& 0.0324	& 2.7247   & $-475$ to +\phn25 & 0.33 & 0.03 & \\
Ni II	& 1741.5531	& 0.0427	& 2.7244   & $-475$ to +\phn25 & 0.30 & 0.02 & \\
Ni II	& 1751.9157	& 0.0277	& 2.7254   & $-475$ to +\phn25 & 0.19 & 0.03 & \\
Zn II	& 2026.1371	& 0.489		& B        & $-475$ to +\phn25 & 0.57\tablenotemark{f} & 0.02 & $W_0$ refers to Zn~II~$\lambda 2026$, Mg~I~$\lambda 2026$ blend. \\*
Zn II	& 2062.6604	& 0.256		& B        & $-475$ to +\phn25 & 0.35 & 0.03 & $W_0$ refers to Cr~II~$\lambda 2062$, Zn~II~$\lambda 2062$ blend. \\*
\enddata
\tablenotetext{a}{Vacuum wavelengths.}
\tablenotetext{b}{Vacuum heliocentric. See note below.}
\tablenotetext{c}{Velocity range for equivalent width measurements relative to $z_{\rm stars}=2.7276$.}
\tablenotetext{d}{Rest frame equivalent width and $1 \sigma$ error.}
\tablenotetext{e}{Blended with stellar C~IV.}
\tablenotetext{f}{Blended with stellar features.}
\end{deluxetable}
\newpage
\vspace*{1cm}
\noindent Notes to Table 1:

\noindent The equivalent width errors listed are from counting statistics
only and were calculated from the S/N of the data across 
each absorption line. Systematic errors due to the uncertainty in the continuum placement
are more difficult to quantify. In general, trials with different continuum fits
showed them to be comparable to the random errors quoted here.
The values of $z_{\rm abs}$ were measured from: 
(i)~the centroid wavelength for unblended lines and 
(ii)~ the wavelength of maximum optical depth for partially blended lines. 
Values in brackets are uncertain for the reasons given in the
`Comments' column.
`B' indicates cases where a reliable value $z_{\rm abs}$
could not be measured because of strong blending.
For blended lines the velocity range given for the
equivalent width measurements refers to {\it each} line in the blend.

\newpage

%
%

\tighten 

\begin{deluxetable}{lllccccc}

\small
\tablewidth{450pt}
\tablecaption{\textsc{Absorption Line Column Densities}\label{tab:columns}}
\tablehead{
  \colhead{} 
& \colhead{} 
& \colhead{} 
& \multicolumn{1}{c}{\phm{gap}} 
& \multicolumn{2}{c}{\phm{99}Optical Depth Method}
& \multicolumn{1}{c}{\phm{gap}} 
& \multicolumn{1}{c}{Profile Fitting\tablenotemark{c}} \\
  \multicolumn{1}{c}{Ion} 
& \multicolumn{1}{c}{$\lambda_{\rm lab}$\tablenotemark{a}} 
& \multicolumn{1}{c}{$f$} 
& \colhead{} 
& \multicolumn{1}{c}{$\Delta v$\tablenotemark{b}} 
& \multicolumn{1}{c}{log $N_{\rm aod}$} 
& \colhead{} 
& \multicolumn{1}{c}{log $N_{\rm pf}$} \\
  \colhead{} 
& \multicolumn{1}{c}{(\AA)} 
& \colhead{} 
& \colhead{} 
& \multicolumn{1}{c}{(km~s$^{-1}$)} 
& \multicolumn{1}{c}{(cm$^{-2}$)} 
& \colhead{} 
& \multicolumn{1}{c}{(cm$^{-2}$)} 
}

\startdata
H I   & 1215.6701  & 0.4164   & & \nodata	      & \phm{$<$}\nodata                 & & \phm{$<$}20.85\\
C I   & 1157.1857  & 0.5495   & & $-475$ to +\phn25 & \phm{$<$}13.55              & & \phm{$<$}13.57\\
N I   & 1134.1653  & 0.01342  & & \nodata 	    & \phm{$<$}\nodata                   & & \phm{$<$}15.35\\
N I   & 1134.4149  & 0.02683  & & \nodata 	    & \phm{$<$}\nodata                   & & \phm{$<$}15.35\\
N I   & 1134.9803  & 0.04023  & & \nodata 	    & \phm{$<$}\nodata                   & & \phm{$<$}15.35\\
Mg II & 1239.9253  & 0.00062  & & \nodata 	    & \phm{$<$}\nodata                   & & \phm{$<$}16.13\\
Mg II & 1240.3947  & 0.00035  & & \nodata 	    & \phm{$<$}\nodata                   & & \phm{$<$}16.13\\
Si II & 1260.4221  & 1.007    & & \nodata	    & \phm{$<$}\nodata                   & & \phm{$<$}15.77\\
Si II & 1304.3702  & 0.094    & & \nodata	    & \phm{$<$}\nodata                   & & \phm{$<$}15.77\\
Si II & 1526.70698 & 0.130    & & \nodata           & \phm{$<$}\nodata                   & & \phm{$<$}15.77\\
Si II & 1808.01288 & 0.00218  & & $-475$ to +\phn25 & \phm{$<$}15.99              & & \phm{$<$}15.77\\
P II  & 1152.818   & 0.2361   & & $-475$ to +\phn25 & $>$14.04\tablenotemark{d}   & & \phm{$<$}14.00\\
S II  & 1250.584   & 0.005453 & & $-475$ to +\phn25 & $>$15.41\tablenotemark{e} & & \phm{$<$}15.43\\
S II  & 1253.811   & 0.01088  & & $-475$ to +\phn25 & \phm{$<$}15.63              & & \phm{$<$}15.43\\
Mn II & 2576.877   & 0.3508   & & $-475$ to +\phn25 & \phm{$<$}13.33              & & \phm{$<$}13.13\\
Fe II & 1144.9379  & 0.106    & & $-775$ to +125    & $>$15.22\tablenotemark{f} & & \phm{$<$}15.00\\
Fe II & 1608.45085 & 0.058    & & $-775$ to +125    & \phm{$<$}15.17              & & \phm{$<$}15.00\\
Fe II & 2344.2130  & 0.114    & & \nodata           & \phm{$<$}\nodata                   & & \phm{$<$}15.00\\
Fe II & 2374.4603  & 0.0313   & & $-775$ to +125    & \phm{$<$}15.25\tablenotemark{g}    & & \phm{$<$}15.00\\
Fe II & 2382.7642  & 0.300    & & \nodata	    & \phm{$<$}\nodata                   & & \phm{$<$}15.00\\
Ni II & 1317.217   & 0.07     & & $-475$ to +\phn25 &	\phm{$<$}14.32            & & \phm{$<$}14.10\\
Ni II & 1370.132   & 0.0769   & & $-475$ to +\phn25 & $<$14.34\tablenotemark{h} & & \phm{$<$}14.10\\
Ni II & 1709.6042  & 0.0324   & & $-475$ to +\phn25 & $<$14.63\tablenotemark{i} & & \phm{$<$}(14.10)\tablenotemark{i}\\
Ni II & 1741.5531  & 0.0427   & & $-475$ to +\phn25 & $<$14.45\tablenotemark{i} & & \phm{$<$}(14.10)\tablenotemark{i}\\
Ni II & 1751.9157  & 0.0277   & & $-475$ to +\phn25 & $<$14.43\tablenotemark{i} & & \phm{$<$}(14.10)\tablenotemark{i}\\
\enddata
\tablenotetext{a}{Vacuum wavelengths.}
\tablenotetext{b}{Velocity range used for the apparent optical depth method.}
\tablenotetext{c}{Using the parameters $b=70{\rm ~km~s}^{-1}$ and $v=-265{\rm ~km~s}^{-1}$, except for H~I (see Figure 3).}
\tablenotetext{d}{Blended in red wing.}
\tablenotetext{e}{Blended in wings.}
\tablenotetext{f}{Blended in blue wing.}
\tablenotetext{g}{Linearly interpolated across bad pixels.}
\tablenotetext{h}{Blended with an intervening Si~IV~$\lambda 1393$ line at $z_{\rm abs} = 2.6606$; the 
contribution of the latter to the blend was assessed with reference to the weaker member
of the Si~IV doublet at $\lambda 1402$.}
\tablenotetext{i}{Blended with stellar features.}
\end{deluxetable}

\newpage

%
%

\begin{deluxetable}{lcccc}
\small
\tablewidth{0pt}
\tablecaption{\textsc{Interstellar Abundances}\label{tab:abundances}}
\tablehead{
\multicolumn{1}{c}{Ion} 
& \multicolumn{1}{c}{log $N$ (cm$^{-2}$)}
& \multicolumn{1}{c}{log (X/H)}
& \multicolumn{1}{c}{log (X/H)$_{\odot}$\tablenotemark{a}}
& \multicolumn{1}{c}{[X/H]$_{\rm cB58}$\tablenotemark{b}}
}

\startdata
H I	& 20.85 & \nodata & \nodata & \nodata \\
N I	& 15.35 & $-5.50$ & $-4.07$ & $-1.43$ \\
Mg II	& 16.13 & $-4.72$ & $-4.42$ & $-0.30$ \\
Si II	& 15.99 & $-4.86$ & $-4.44$ & $-0.42$ \\
P II    & 14.20\tablenotemark{c} & $-6.65$ & $-6.44$ & $-0.21$ \\
S II	& 15.63 & $-5.22$ & $-4.80$ & $-0.42$ \\
Mn II   & 13.33 & $-7.52$ & $-6.47$ & $-1.05$ \\
Fe II   & 15.20 & $-5.65$ & $-4.50$ & $-1.15$ \\
Ni II	& 14.32 & $-6.53$ & $-5.75$ & $-0.78$ \\

\enddata
\tablenotetext{a}{Solar meteoritic abundance scale from Grevesse \& Sauval (1998),
except N for which we adopt the recent reevaluation of the solar photospheric abundance 
by Holweger (2001).}
\tablenotetext{b}{[X/H]$_{\rm cB58}$ = log (X/H) $-$ log (X/H)$_{\odot}$.}
\tablenotetext{c}{Corrected by 0.2 dex (see text in \S5.1).}
\end{deluxetable}

\newpage

%
%

\begin{figure}
\figurenum{1}
\psfig{figure=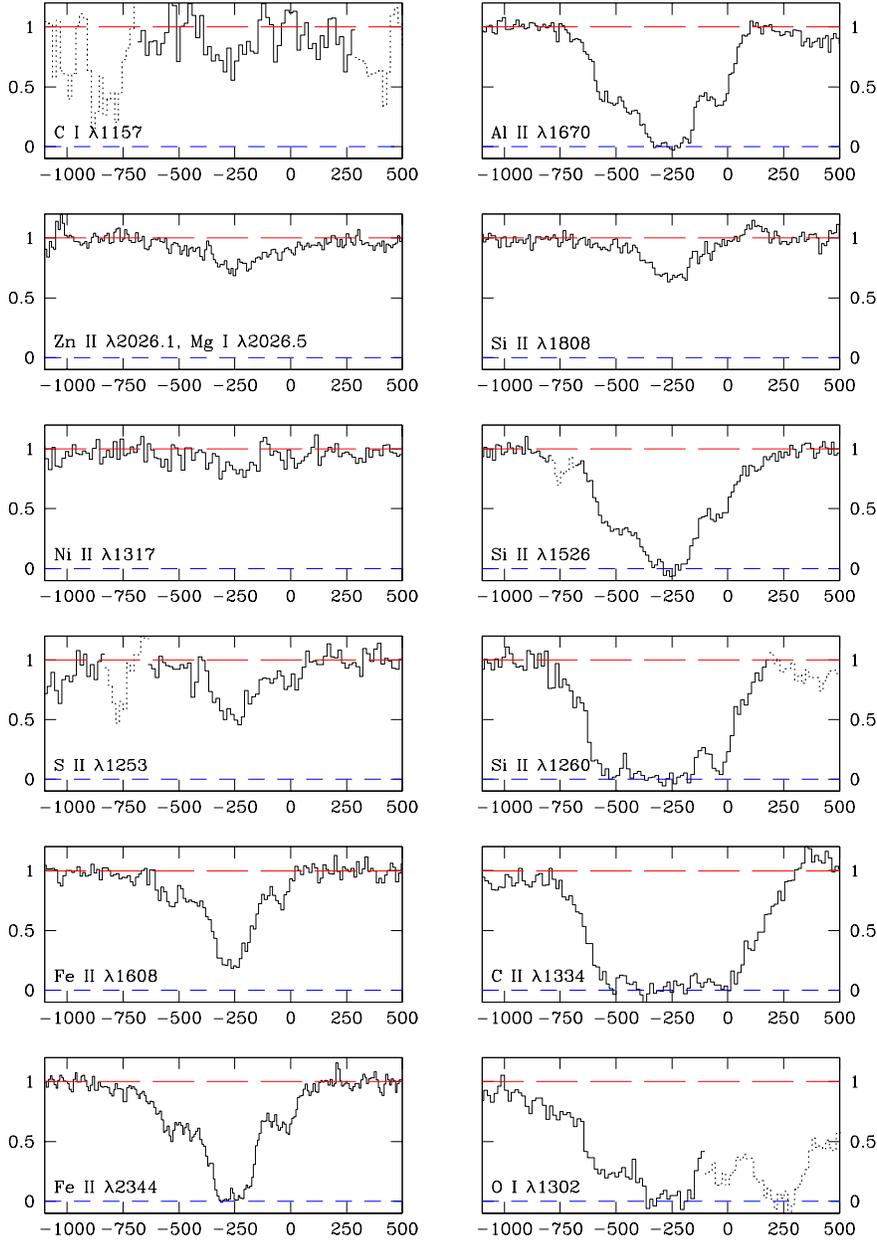,width=165mm}
\vspace{0.25cm}
\figcaption{Normalised profiles of 
interstellar absorption lines in cB58. In this figure
we have collected examples of absorption lines from species which are 
the dominant ion stages in H~I regions (except for C~I), 
illustrating a range of
line strengths. The $y$-axis is residual intensity, while
the $x$-axis is velocity (in km~s$^{-1}$) relative
to the stellar systemic redshift $z_{\rm stars} = 2.7276$ (see 
\S 3.1). Each bin corresponds to a wavelength interval 
$\delta \lambda = 0.067$\,\AA, equivalent to a velocity
interval $\delta v = 8.5$--17.5\,km~s$^{-1}$ for the 
transitions shown here. Dotted lines are used to 
indicate blends with absorption
features which are {\it not} those labelled in each panel. 
} 
\end{figure}

%
%

\begin{figure}
\figurenum{2}
\psfig{figure=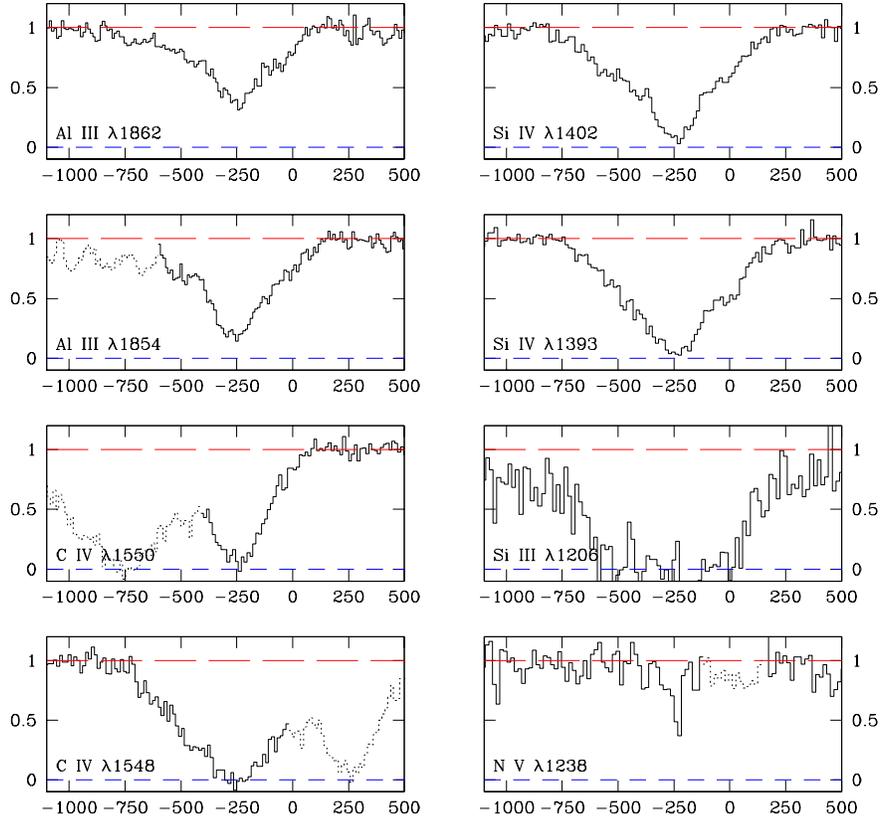,width=165mm}
\vspace{-3.5cm}
\figcaption{Same as Figure 1, showing interstellar absorption line profiles
of highly ionized species covered by our ESI spectrum of cB58. }
\end{figure}

%
%

\begin{figure}
\figurenum{3}
\vspace*{-3.5cm}
\hspace*{-2.0cm}
\psfig{figure=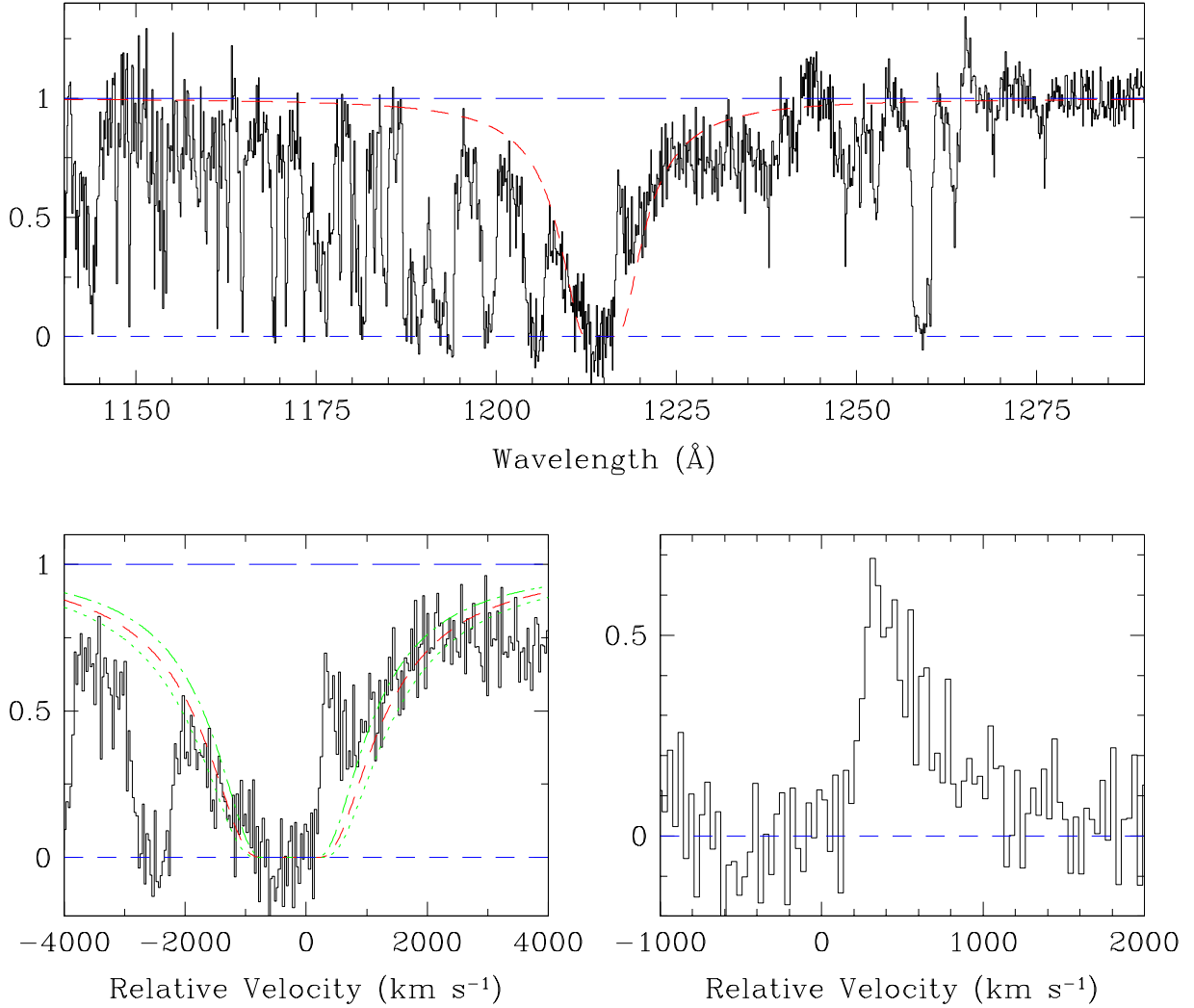,width=195mm}
\vspace{-7.5cm}
\figcaption{The \lya\ profile in cB58. {\it Top 
panel}: Black histogram, observed spectrum; dashed line, theoretical 
absorption profile for $N$(H~I)$ = 7.0 \times 10^{20}$~cm$^{-2}$. 
{\it Bottom left-hand panel}: Same as the top panel, but on a 
velocity scale relative to $z_{\rm stars} = 2.7276$. 
Also shown are damped profiles for $N$(H~I)$ = 5.5 \times 10^{20}$
and $8.5 \times 10^{20}$~cm$^{-2}$.
{\it Bottom right-hand panel}: Residual \lya\ emission after subtraction of
the $N$(H~I)$ = 7.0 \times 10^{20}$~cm$^{-2}$ absorption component.
The $y$-axes of the plots show residual 
intensity; each bin corresponds to a wavelength interval 
$\delta \lambda = 0.134$\,\AA\
($\delta v = 33$\kms).
}
\end{figure}

%
%

\begin{figure}
\figurenum{4}
\vspace*{-1cm}
\hspace*{-0.55cm}
\psfig{figure=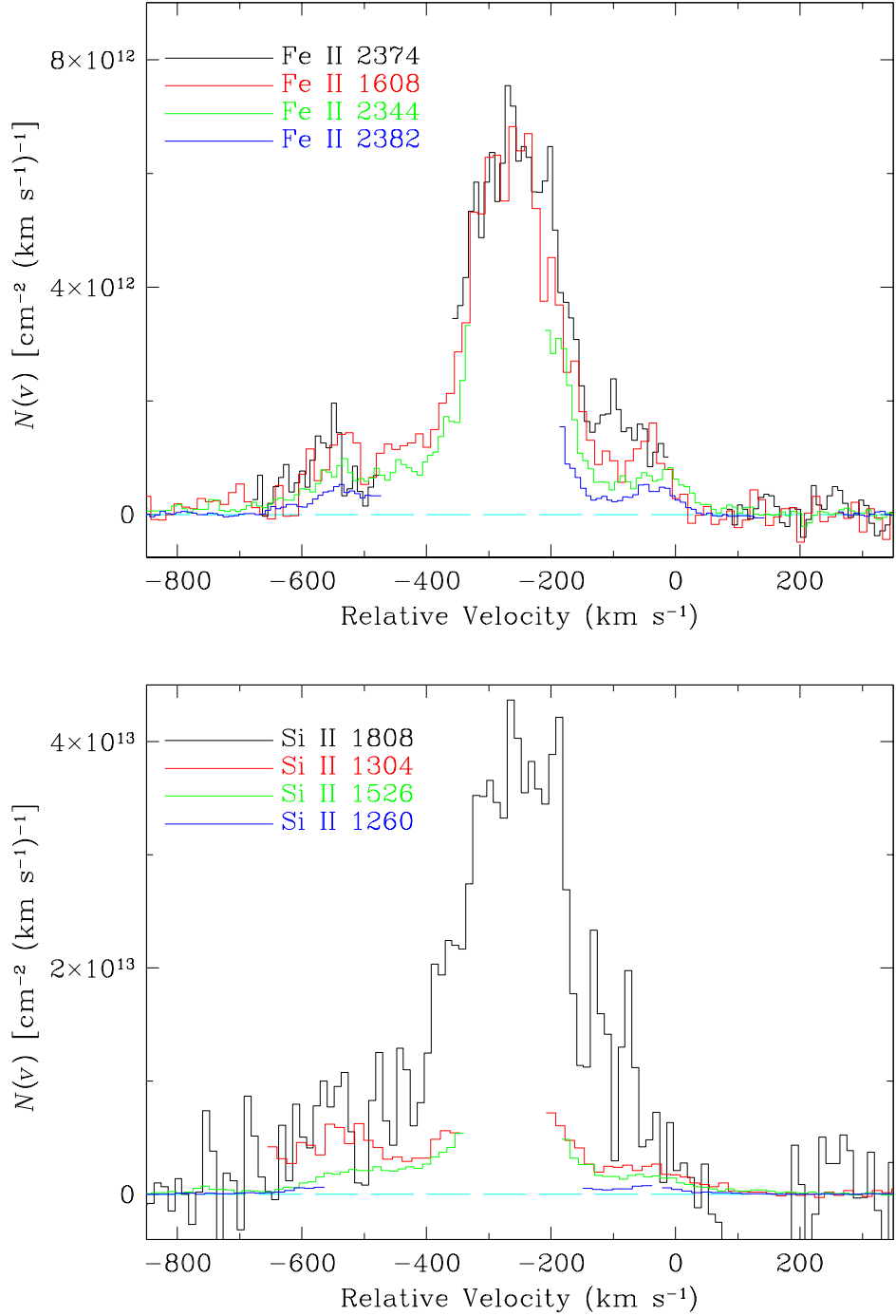,width=180mm}
\vspace{-2.5cm}
\figcaption{Apparent optical depth method: Column density as a function of velocity
relative to $z_{\rm stars} = 2.7276$ for Fe~II and Si~II lines. The transitions
are shown in increasing order of $f \lambda$ starting from the top.}
\end{figure}

%
%

\begin{figure}
\figurenum{5}
\vspace*{1cm}
\hspace*{-0.55cm}
\psfig{figure=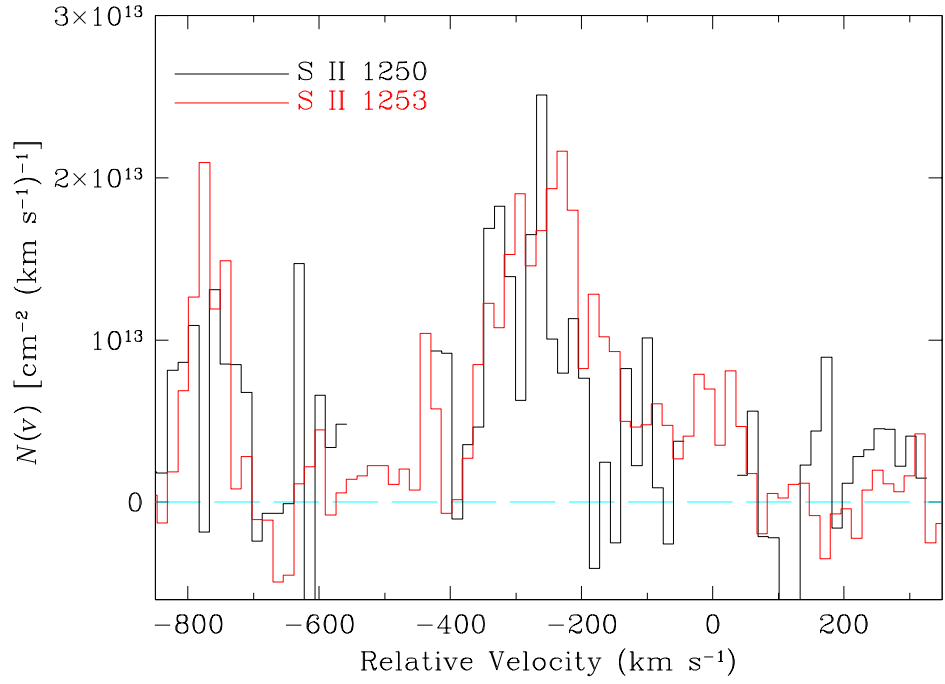,width=180mm}
\vspace{-10.5cm}
\figcaption{Apparent optical depth method: Column density as a function of velocity
relative to $z_{\rm stars} = 2.7276$ for two lines of the triplet
S~II~$\lambda \lambda 1250, 1253, 1259$. The products $f \lambda$ 
for S~II~$\lambda 1250$ and $\lambda 1253$ respectively 
are in the ratio 1:2\,.
The third line, S~II~$\lambda 1259$,
is blended with Si~II~$\lambda 1260$.}
\end{figure}

%
%

\begin{figure}
\figurenum{6}
\vspace*{-1cm}
\hspace*{-0.55cm}
\psfig{figure=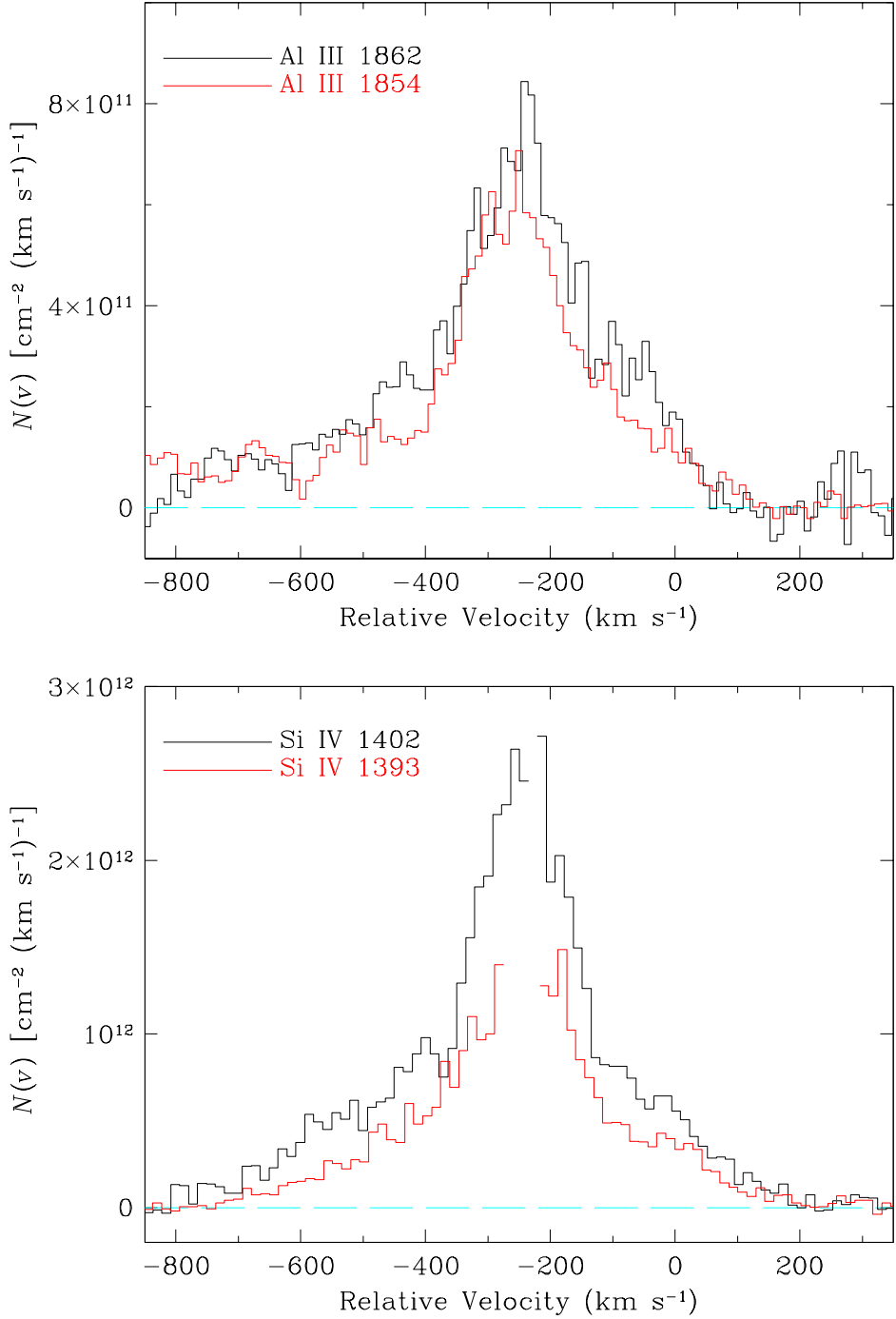,width=180mm}
\vspace{-2.5cm}
\figcaption{Apparent optical depth method: Column density as a function of velocity
relative to $z_{\rm stars} = 2.7276$ for the Al~III and Si~IV doublets. In each
case the line at the shorter wavelength is the stronger member of the doublet.}
\end{figure}

%
%

\begin{figure}
\figurenum{7}
\hspace*{1.25cm}
\psfig{figure=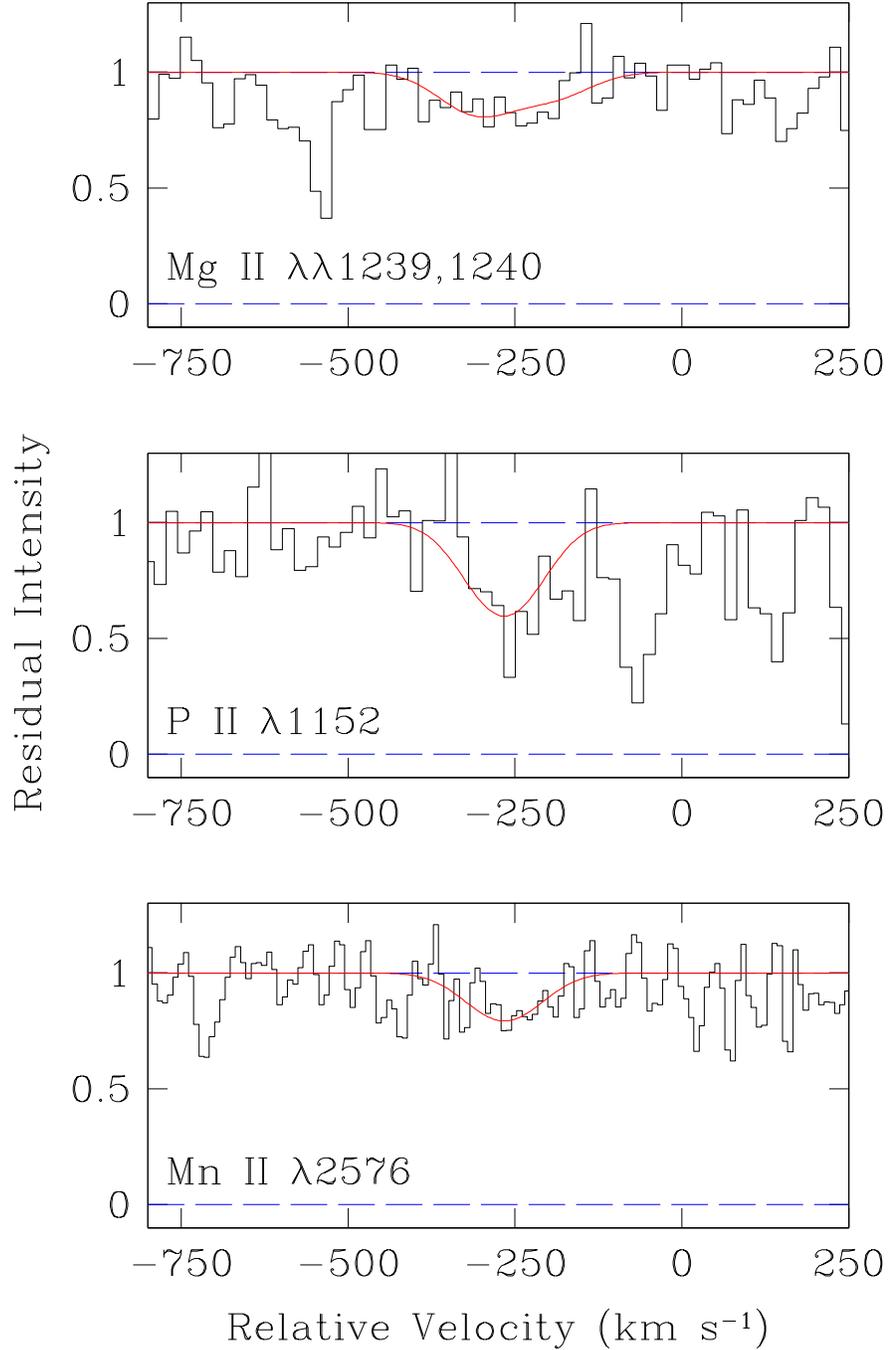,width=145mm}
\figcaption{Examples of profile fits to weak absorption lines.
The fits are for a single absorption component 
with velocity dispersion parameter $b = 70$\kms\
centered at $v = -265$\kms\ and convolved with the
instrumental broadening profile 
which is a Gaussian of FWHM = 58\,\kms.
The velocity scale is relative to $z_{\rm stars} = 2.7276$\,.}
\end{figure}

%
%

\begin{figure}
\figurenum{8}
\vspace*{-3cm}
\hspace*{-0.55cm}
\psfig{figure=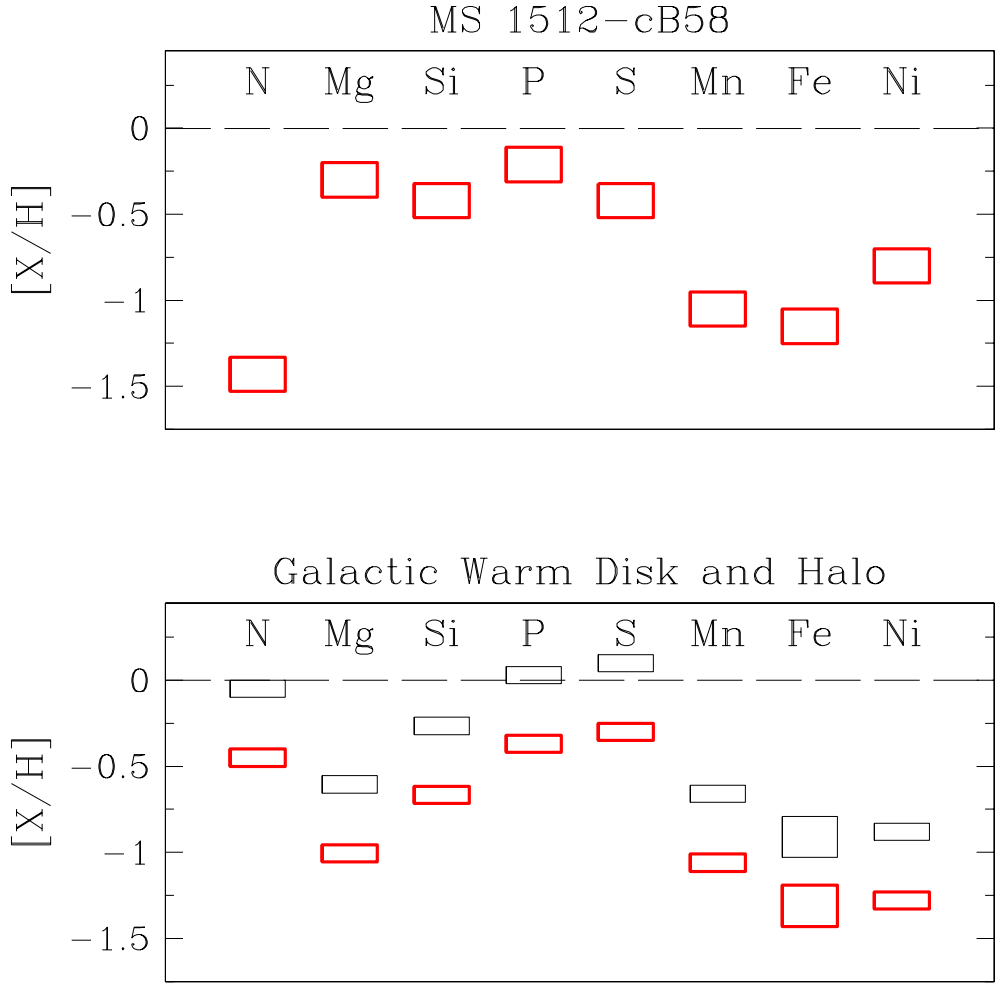,width=180mm}
\vspace{-7.5cm}
\figcaption{{\it Top Panel:} Element abundances in the interstellar gas of cB58.
The height of the boxes reflects the $\pm 0.1$\,dex statistical uncertainty
in the ion column densities; sources of systematic error discussed in the text 
are expected to be less than about 0.2\,dex.~ {\it Lower Panel:} The thin
boxes show the abundances of the same elements measured in diffuse clouds 
of the Milky Way (see text for references). 
The height of the box for Fe
reflects the range of abundances among different sight-lines which sample
both disk and halo clouds; the same range probably applies to other elements
but, in the absence of direct measurements, these boxes have been set to 
$\pm 0.05$\,dex. The thick boxes show the same abundance pattern shifted
down by 0.4\,dex, so as to facilitate the comparison with cB58.}
\end{figure}

%
%

\begin{figure}
\figurenum{9}
\hspace*{-0.55cm}
\psfig{figure=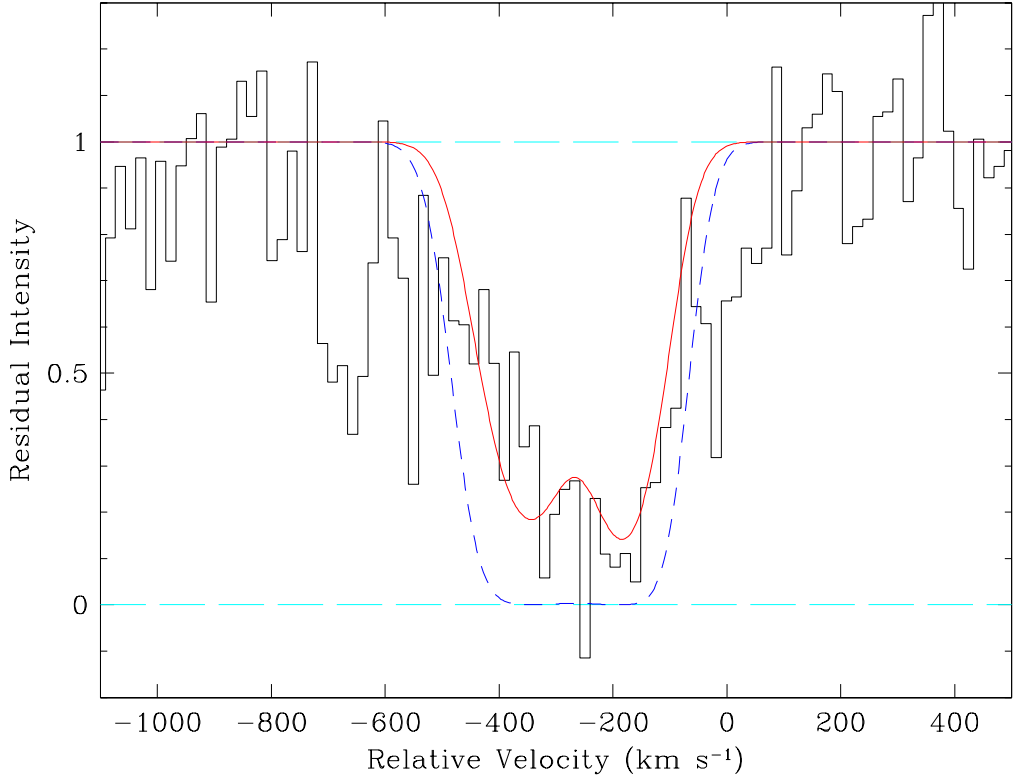,width=180mm}
\vspace{-10.5cm}
\figcaption{The blended N I $\lambda$1134 triplet in cB58. 
The observed profile (histogram) is shown on a velocity scale relative to $z_{\rm stars}$. 
Also shown are theoretical absorption profiles for the
main component of the absorption lines, a single `cloud' 
with velocity dispersion parameter
$b = 70$\,\kms\ centered at $v = -265$\,\kms.
The continuous line is our best fit, given by a nitrogen column density
$N{\rm (N~I)}=2.2\times 10^{15}~{\rm cm}^{-2}$. The dashed line shows the 
absorption profile corresponding to the nitrogen abundance
deduced by Teplitz et al. (2000), which is clearly inconsistent
with our data.}
\end{figure}

\end{document}